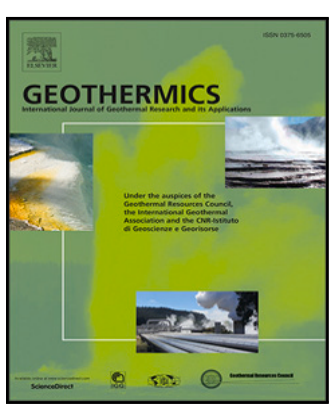

# Predicting the geothermal gradient in Colombia: A machine learning approach

Juan C. Mejía-Fragoso *, Manuel A. Flórez, Rocío Bernal-Olaya

*Universidad Industrial de Santander, Carrera 27 Calle 9, Bucaramanga, 680002, Santander, Colombia*

## ARTICLE INFO

## ABSTRACT

Accurately determining the geothermal gradient is crucial for assessing geothermal energy potential. In Colombia, despite an abundance of theoretical geothermal resources, large regions of the country lack gradient measurements. This study introduces a machine learning approach to estimate the geothermal gradient in regions where only global-scale geophysical datasets and course geological knowledge are available. We find that a Gradient-Boosted Regression Tree algorithm yields optimal predictions and extensively validates the trained model, obtaining predictions of our model within 12% accuracy. Finally, we present a geothermal gradient map of Colombia that serve as an indicator of potential regions for further exploration and data collection. This map displays gradient values ranging from 16.75 to 41.20 °C/km and shows significant agreement with geological indicators of geothermal activity, such as faults and thermal manifestations. Additionally, our results are consistent with independent findings from other researchers in specific regions, which supports the reliability of our approach.

## 1. Introduction

The tectonic and geological characteristics of Colombia give rise to thermal anomalies and regions exhibiting elevated geothermal gradients, which represent promising sources of geothermal energy. However, performing accurate estimations of the country's potential for geothermal energy generation remains challenging. The most current geothermal gradient map for Colombia covers roughly half of the country's territory (Alfaro et al., 2009), for the rest no systematic determinations of the geothermal gradient are available. Some studies have tried to address this issue. Uyeda and Watanabe (1970) observed normal to subnormal geothermal gradient values across South America, with higher values in the Andes related to geothermal activity. Bachu et al. (1995) focused on the Llanos Basin, finding that geothermal gradients decrease with depth and westward. In northwestern Colombia, Quintero et al. (2019) estimated geothermal gradients using grid systems and aeromagnetic data. Matiz-León (2023) emphasized the importance of statistical methods and spatial prediction for estimating geothermal gradients in the Llanos Basin, especially in areas lacking in-situ data. Unfortunately, all of these studies are either focused on narrow areas of interest or scales that are too large for meaningful and detailed geothermal energy estimations. Measurements from boreholes drilled by the oil and gas industry provide the best direct constraint on the geothermal gradient; in the case of Colombia, they are either very sparse or clustered to specific regions, as the need to decarbonize the economy grows, it is unlikely that the past pace of oil exploration activity will continue. Therefore, a different approach is needed if Colombia is to realize its clean energy production objectives (Alfaro et al., 2021).

It is seldom impossible to develop comprehensive geothermal gradient models for an entire country using only direct borehole measurements (Burton-Johnson et al., 2020), so indirect approaches are necessary. Geophysical methods provide complementary estimates of the geothermal gradient over large spatial scales and depths that are inaccessible by boreholes alone; magnetic and electrical surveys exploit the thermal properties of rocks to estimate temperature gradients. While ideal, due to their cost, such surveys are typically performed once a narrow target of interest has been identified. Numerical thermal modeling has emerged as a useful tool for estimating geothermal gradients as it allows for the prediction of the entire temperature field of the region's subsurface (Békési et al., 2020). Thermal models require detailed knowledge of the underlying geology, direct temperature and heat flow measurements and involve a significant number of assumptions. For Colombia, while in principle modeling could be performed, it is unlikely to yield accurate results.

Data-driven approaches have become viable alternatives for the estimation of the earth's thermal properties. Goutorbe et al. (2011) explored empirical estimators to estimate heat flow in data-scarce regions, using global geological and geophysical proxies; obtaining results that demonstrate their effectiveness in approximating heat flow. Lucazeau (2019) updates the global terrestrial heat flow dataset with around 70,000 measurements, finding that Earth's total heat loss is approximately 40–42 TW, closely matching previous conductive cooling estimates. The study highlights improved measurement quality and coverage, especially in young oceanic regions, which significantly reduce discrepancies with existing models.

* Corresponding author.
*E-mail addresses:* juan2228283@correo.uis.edu.co (J.C. Mejía-Fragoso), maflotor@uis.edu.co (M.A. Flórez), rbernalo@uis.edu.co (R. Bernal-Olaya).

Rezvanbehbahani et al. (2017) used a gradient boosting regression tree algorithm and a catalog of global geological features to produce a new geothermal heat flow (GHF) map of Greenland, their predictions were consistent with regional geology, tectonics and ice core measurements. A similar approach was used to predict GHF in Antarctica by Lösing and Ebbing (2021), it incorporated regional datasets to improve predictions. Advances in machine learning and the increasing availability of regional and global geological datasets allow for the development of cost effective approaches to provide initial baselines of the thermal properties of entire countries.

Machine learning has increasingly been used in many geothermal-related studies as well. Assouline et al. (2019) proposed a methodology to extract the very shallow geothermal potential at a national scale for Switzerland, using a combination of geographic information systems, traditional modeling, and machine learning; this approach leverages data such as monthly temperatures at various depths in the surface layer, along with thermal conductivity and diffusivity. Siler et al. (2021) applied unsupervised machine learning to identify key geological factors influencing geothermal production in Nevada's Brady geothermal field, by analyzing geological data to discern controls on geothermal fluid pathways. Additionally, Pang et al. (2023) optimized machine learning algorithms to predict thermal conductivity (TC) using accessible, high-resolution logging data, employing techniques including random forest (RF), convolutional neural network (CNN), support vector regression (SVR), and particle swarm optimization-enhanced SVR (PSO_SVR). Furthermore, Alqahtani et al. (2023) leveraged borehole temperature and remote sensing data to pinpoint prospective zones with significant geothermal activity within the Harrat Rahat volcanic field in western Saudi Arabia, favoring exploration and drilling efforts.

In this work we present a Gradient Boosted Regression Tree algorithm to accurately predict the spatial distribution of geothermal gradients across a region. The approach only requires global scale geological and geophysical datasets, coarse geological knowledge and sparsely distributed borehole temperature measurements to produce accurate results. The paper is organized as follows: In Section 3 we describe the details of the algorithm, in Section 4 we present the data and the strategy used to train the model, in Section 5 we provide extensive validation of our model and present a new map of geothermal gradients for Colombia. The predicted map highlights regions displaying relatively high geothermal gradients, where future exploration surveys should be conducted. Further validation of the model is provided by the fact that the predictions show good agreement with indirect geophysical measurements of the geothermal gradient performed in the Amazon.

## 2. Geological setting of Colombia

Colombia, located in northwestern South America, and positioned at the intersection of the Nazca, South American, and Caribbean plates, has a dynamic geological evolution extending back to 1780 million years. This dynamism is characterized by the Andean orogeny, which started in the Jurassic and is still active today, along with significant faulting and seismic activity. The Guiana Shield is part of the ancient continental core of northern South América, it consists of igneous intrusive rocks, volcanic rocks, and metamorphic rocks, such as granites, trachytes, and gneisses from the Precambrian era. The oldest rocks dated in Colombia are part of the Migmatitic Complex near San Felipe in Guainía, at the southeast of the Eastern Llanos, with an age of 1780 million years. Similar rocks are found in several other regions in Colombia, indicating the expansive reach of the Guiana Shield across northern South America (Lobo, 1987).

The collision of the Panama Block, a segment of the Caribbean Plate, with South America in the Pliocene epoch was the event that joined the Americas. The tectonic interactions involving this block, including the collision with the South American plate, have been instrumental in shaping the Isthmus of Panama and the closure of the Central American Seaway, which had profound impacts on ocean circulation and climate. (de Porta, 2003).

The early stages of the Andean orogeny during the Jurassic and Cretaceous were marked by extensional tectonics, including rift and back-arc basin development, and the emplacement of large batholiths, associated with the subduction of cold oceanic lithosphere. In the mid to late Cretaceous (around 90 Ma), the orogenic character shifted abruptly as younger and hotter lithosphere began subducting beneath South America. This led to intense compressional deformation, Andean uplift, and subsequent erosion from the late Cretaceous onward (Ramos, 2009). The Colombian Andes are divided in three (Lobo, 1987; Gómez-Tapias et al., 2020) (Fig. 1):

- **The Central Cordillera (CC)** is constituted by a complex geology with a diversity of rock types, igneous rocks are predominant (including the Antioquian Batholith), but metamorphic and sedimentary rocks are also present. It was formed through intense tectonic activities, including the uplifting and folding of sedimentary deposits, and is significant for its mineral wealth, particularly gold and silver. The CC features a Triassic polymetamorphic basement, intruded by plutons from the Permian to the Cenozoic, linked to the subduction of the Nazca Plate. This structure is flanked by Mesozoic intrusions and Jurassic volcaniclastic sequences, as well as exposed high-grade metamorphic rocks from previous orogenies. Cretaceous marine and volcanoclastic rocks, along with Neogene to Quaternary volcanoes, some still active, characterize the CC. The region also includes Miocene deposits and is bordered by major faults that define its tectonic limits.
- **The Eastern Cordillera (EC)** is the broadest of the three and is noted for its rich deposits of coal and oil, which are the result of a sedimentary history that includes shallow marine environments from past geologic periods. Its complex geological history includes significant episodes of tectonic folding and faulting. The EC basement consists of high-grade metamorphic rocks from the Mesoproterozoic to Neoproterozoic eras, overlaid by Ordovician low-grade metamorphic rocks and Paleozoic sedimentary sequences. These are further topped with a thick succession of Cretaceous marine and Cenozoic continental sedimentary rocks, all deformed during the Andean orogeny. The region also features Jurassic sedimentary, intrusive, and volcaniclastic sequences. Its boundaries are delineated by major fault systems that mark the transition to Caguán–Putumayo and Llanos Orientales Basins (Fig. 2).
- **The Western Cordillera (WC)** is the lowest and shortest of the three and is marked by significant volcanic activity. This cordillera serves as a barrier between the Pacific Ocean and the rest of Colombia. It is delimited by the Cauca River valley to the east, which separates it from the Central Cordillera. The WC primarily comprises Cretaceous sedimentary rocks, gabbros, and basalts from the Caribbean-Colombian oceanic plateau, which were accreted to Colombia's western margin from the Late Cretaceous to the Paleogene period. In its southern region, the Cordillera hosts Paleogene plutonic and volcanoclastic rocks. Towards the north, it features Miocene basalts and Pliocene volcanoclastic rocks, interspersed with minor Neogene intrusions. The range's southern extremity contains Neogene and Quaternary volcanic deposits, with several active volcanoes that constitute the Southern Volcanic Segment of Colombia.

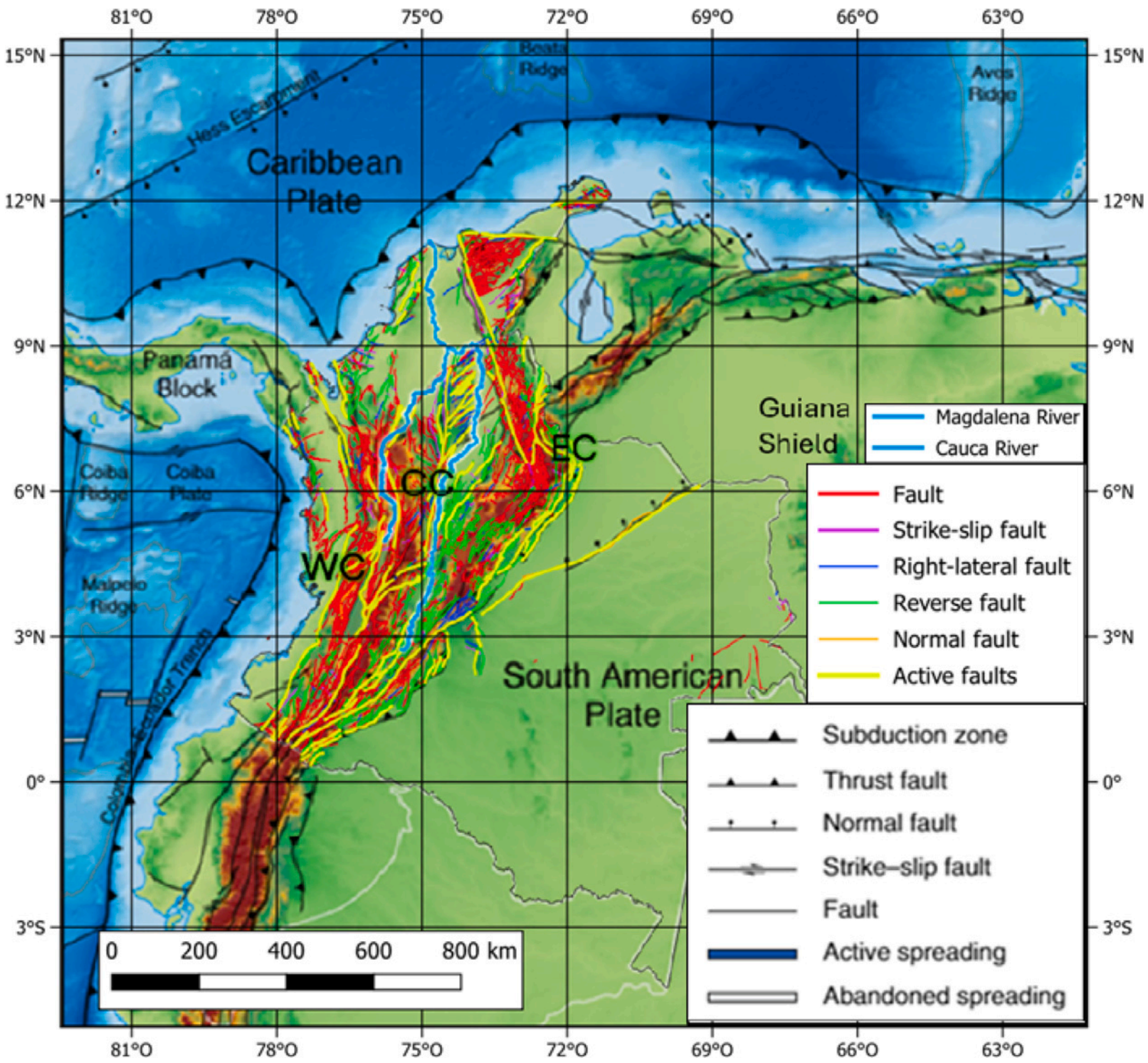

**Fig. 1.** Tectonic scheme of Colombia, showing the Eastern Cordillera (EC), Central Cordillera (CC), Western Cordillera (WC); the different types of faults, the active faults, the Magdalena and Cauca rivers. Modified from Gómez-Tapias et al. (2020).

CC and EC are separated by the Magdalena River Valley, which runs roughly parallel to the length of the two cordilleras and is a key feature of Colombia's geography, hosting one of the major rivers in the country, the Magdalena River. The Middle Magdalena Valley Basin (Fig. 2), part of the Magdalena River Valley, is the most extensively explored in Colombia and remains one of the most prolific areas as it hosts significant hydrocarbon resources. More than forty oil fields have been discovered within Cenozoic sediments, including Colombia's first giant oil field, La Cira-Infantas. However, Cretaceous targets in carbonates remain underexplored (ANH (Agencia Nacional de Hidrocarburos), 2010a). The geological sequence in the basin begins with Jurassic continental deposits followed by Cretaceous sediments that transition from calcareous to siliciclastic, indicating a shift from transitional to marine environments. The Paleogene sequence consists predominantly of siliciclastic rocks deposited under mostly continental conditions with some marine influences. The basin has undergone three major deformational phases – rifting, thrusting, and wrenching – that have shaped various trap geometries (Barrero et al., 2007).

The Upper Magdalena Basin, which is also part of the Magdalena River Valley, is a Neogene Broken Foreland Basin, evolved from a larger Paleogene foreland basin extending eastward towards the Guyana Shield, covering an area of approximately 26,200 $km^2$. It is flanked by Precambrian to Jurassic basement uplifts of the Eastern and Central Cordilleras. The basin's stratigraphy begins with a Cretaceous continental sequence, followed by transgressive shales and limestones, and the Caballos sandstone unit, a key hydrocarbon reservoir. Subsequent layers include Albian to Campanian limestones, shales, and cherts, with middle Albian and Turonian organic-rich source rocks fueling all hydrocarbon accumulations. The sequence culminates with the Campanian to Maastrichtian Monserrate Formation, primarily sandstone and another significant hydrocarbon reservoir. The post-collisional Cenozoic sequence, composed of Paleogene and Neogene molasses, is entirely non-marine (Barrero et al., 2007).

The separation between the CC and the WC is known as the Cauca River Valley, it also runs parallel to the cordilleras and contains the Cauca River, which is the second major river in Colombia. It is narrower compared to the Magdalena River Valley and is flanked by volcanic peaks from the Central Cordillera. The Cauca River Valley is also noted for its gold mining, given the historical presence of gold in the surrounding geological formations, and it also has significant hydroelectric potential due to the flow of the river, supporting multiple hydroelectric plants that contribute to the energy needs of the country (Lobo, 1987).

The Llanos Basin, another significant region located in Eastern Colombia, is bounded by the Colombian-Venezuelan border to the north, the Guaicáramo fault system to the west, and the Guyana Shield to the east and the south. Its geological development began in the Paleozoic with a rifting phase, depositing siliciclastic sediments over a crystalline Precambrian basement. From the Triassic through the Late Cretaceous, it served as the eastern shoulder of a major rift system, then transitioned into a foreland basin from the Maastrichtian to the Paleocene. It has since collected thick molasse deposits up to the present (Barrero et al., 2007). The basin contains over 1700 MMBO of documented recoverable oil, including two giant fields, three major fields, and over eighty minor fields (ANH (Agencia Nacional de Hidrocarburos), 2010b).

The Caguán-Putumayo Basin, spanning 110,304 $km^2$ across southwestern Colombia, shares its geological history with Ecuador's Oriente Basin as part of a foreland basin system. It is bounded by the Eastern Cordillera to the west, the Guyana Shield to the east, the Serranía de la Macarena to the north, and the international border with Ecuador to the south. The exploration of the basin has focused on structural traps involving Cretaceous and Cenozoic formations, with significant hydrocarbon prospects. The northern part of the basin, less explored, shows evidence of at least one active oil system, with structural features such as high-angle reverse faults, associated folds, and blind faults in the foothills, alongside normal faults and wedging in the foreland zone (Barrero et al., 2007; ANH (Agencia Nacional de Hidrocarburos), 2010a).

The Vaupés-Amazonas Basin is a southeast-plunging depression located in southeastern Colombia. It comprises Lower Paleozoic rocks forming structural highs. An isolated area north of Mitú, covered by Neogene sediments, is also included within this basin. Extending south-southeast, the basin reaches the borders of Peru and Brazil, potentially connecting with the Solimoes Basin further south (Barrero et al., 2007).

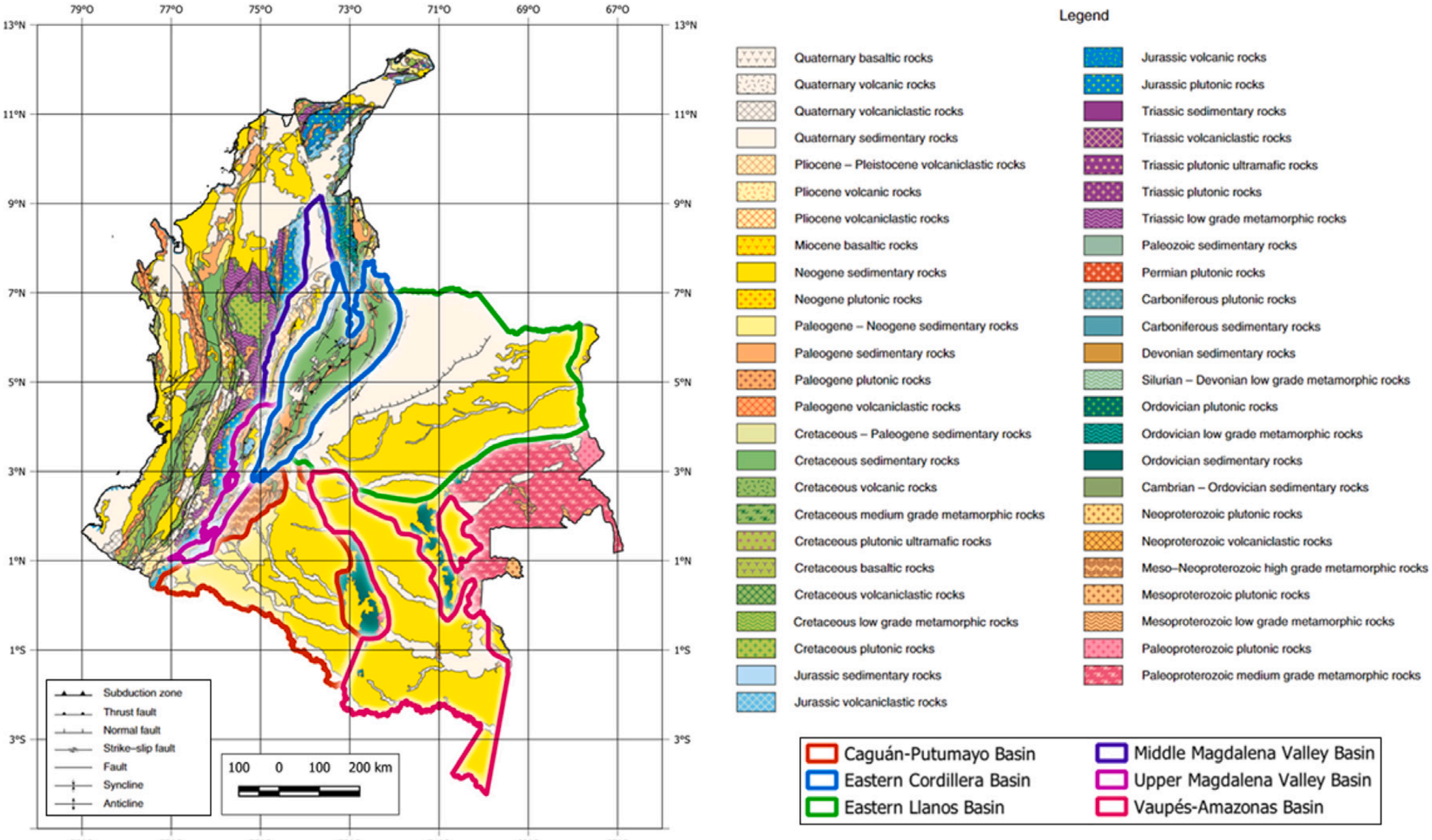

**Fig. 2.** Geological map of Colombia highlighting the main basins. Modified from Gómez-Tapias et al. (2020).

### 2.1. Geothermal gradient in Colombia

The scientific study of Earth's heat has practical applications in exploiting heat as an energy source (both thermal and electric), as well as in understanding hydrocarbon and mineral resources, and modeling the evolution of Earth's crust and tectonic processes (Alfaro et al., 2009). Colombia, located within the Pacific Ring of Fire – a tectonic belt renowned for its volcanoes and earthquakes – hosts regions that are highly favorable for generating geothermal energy. The country is home to 13 active volcanoes and numerous hot springs (Fig. 3), highlighting its substantial geothermal potential (Clavijo et al., 2008; Marzolf, 2014).

Alfaro et al. (2009) defined a preliminary geothermal gradient map to promote research on terrestrial heat flow and produce a map based on measured information. For the preliminary map, apparent geothermal gradients were calculated from temperature measurements at the maximum depth recorded (BHT) in approximately 4600 wells, primarily drilled by the oil sector, that were adjusted using the AAPG's empirical method and included average surface temperatures. The wells, spread across 14 sedimentary basins, cover about 50% of the country's territory (Fig. 4a)

More contemporary and methodologically varied studies were conducted to estimate the geothermal gradient in more localized regions, such as the one by Matiz-León (2023) for the Eastern-Llanos. To calculate the geothermal gradient, they utilized BHT measurements, applied deterministic methods such as Minimum Curvature and Inverse Weighted Distance (IDW) for initial spatial predictions, and enhanced these with probabilistic methods like Ordinary Kriging and Sequential Gaussian Simulation (SGS) for detailed 2D and 3D spatial modeling. (Fig. 4c).

Gomez et al. (2019) calculated the geothermal gradient for northwestern Colombia (a region characterized by data scarcity) by first estimating the Curie point depth (CPD) using aeromagnetic data. This involved the application of spectral analysis techniques to determine the depth at which magnetic minerals lose their permanent magnetic properties due to temperature increases. The CPD was then correlated with surface heat flow measurements to derive the geothermal gradient across the region (Fig. 4b).

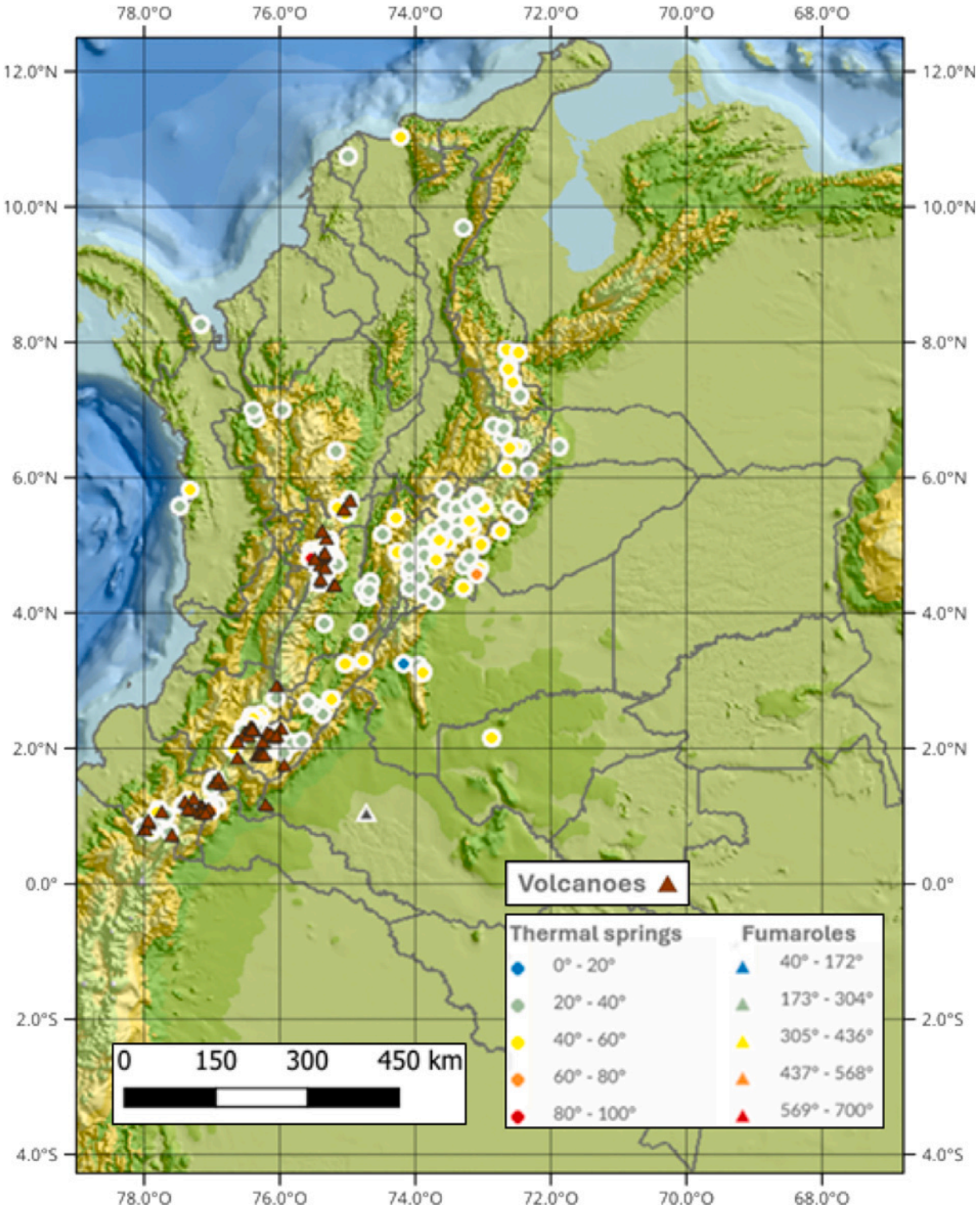

**Fig. 3.** Thermal manifestations and volcanoes in Colombia. Modified from SGC (Servicio Geológico Colombiano) (2015).

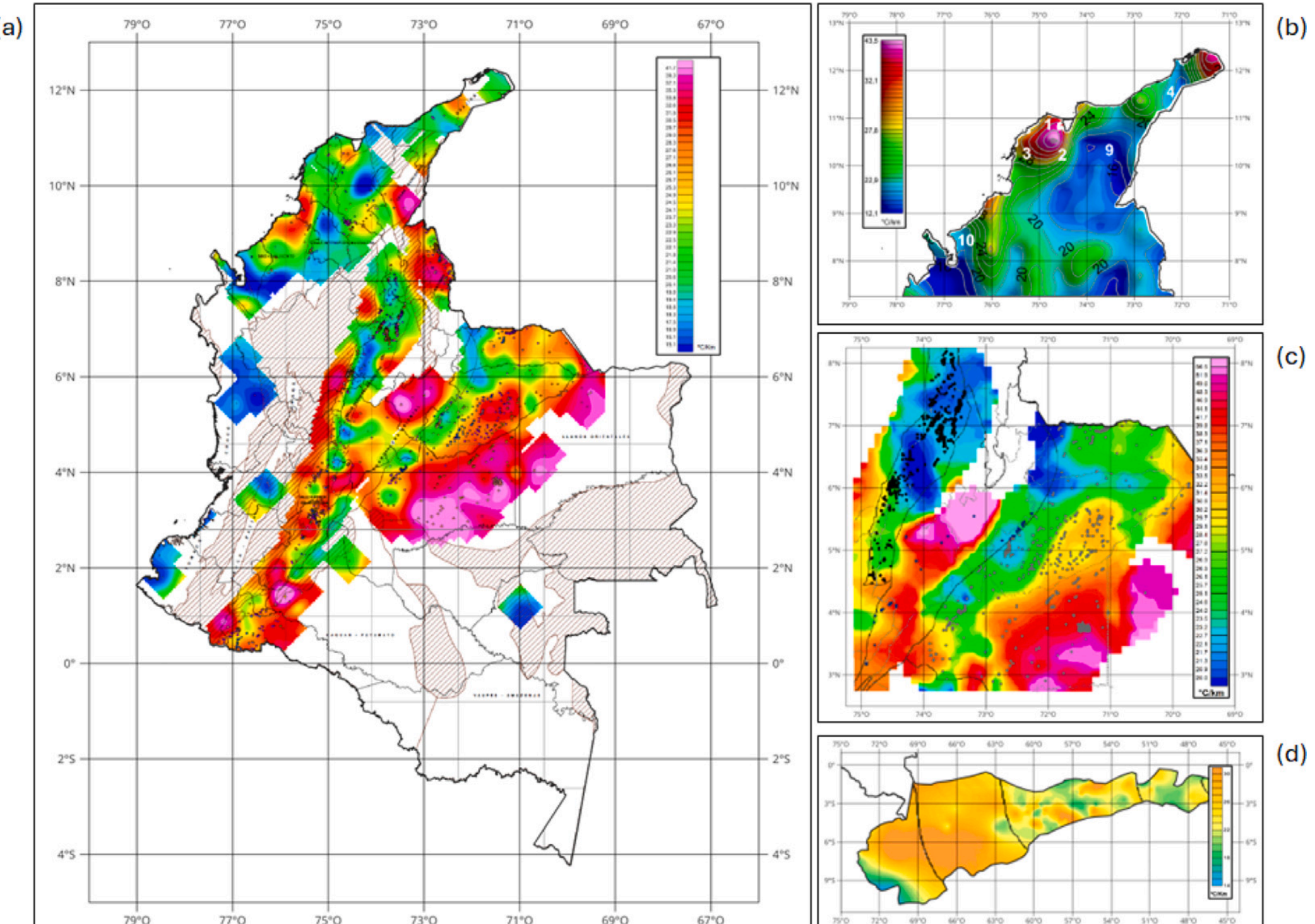

**Fig. 4.** (a) Preliminary geothermal gradient map of Colombia. modified from Alfaro et al. (2009). (b) Colombian-Caribbean Geothermal Gradient Map. modified from Gomez et al. (2019). (c) Geothermal gradient for the Eastern-Llanos basin. modified from Matiz-León (2023). (c) Geothermal gradient of the Amazon basin in northern Brazil. modified from Pimentel and Hamza (2010).

Unfortunately, there are no studies specifically focused on calculating the geothermal gradient in the Colombian Amazon region. However, the Brazilian portion of the Amazon has been studied by Pimentel and Hamza (2010), who determined the geothermal gradients primarily using corrected Bottom Hole Temperature (BHT) data. Their study also employed Constant Bottom Hole Temperature (CBT) methods for more stable thermal fields and Conventional Logging (CVL) to calculate gradients from litological profiles (Fig. 4d).

## 3. Method

To achieve precise quantitative predictions within the geothermal context, it is necessary to utilize a regression algorithm that can effectively capture the relationship between the Geothermal Gradient and its geological setting. Our approach utilizes gradient boosting regression, a technique within the domain of supervised machine learning. This method systematically develops an ensemble of regression trees, where each tree is designed to incrementally adjust to the gradient of the loss function from the previous one, following the methodology outlined by Friedman (2001).

Regression trees are particularly adept at identifying complex nonlinear relationships by recursively partitioning data based on specific criteria, such as maximum depth, number of leaves, and minimum samples per leaf. In the gradient-boosting framework, each iteration involves training a new tree on a subset of the data to accurately predict gradient descent steps. By aggregating these trees into an ensemble model, we improve our ability to predict outcomes for new data samples. This method stands out for its straightforward interpretability and operational simplicity, qualities that are often highlighted in the literature (Lösing and Ebbing, 2021).

We considered several machine learning algorithms, including Random Forest Regression. Our decision to utilize the Gradient Boosted Regression Tree algorithm was primarily influenced by its superior performance in terms of error metrics when compared to alternatives.

Specifically, the eXtreme Gradient Boosted Regression Tree, or XGBoost, operates on a sequential model building process where each successive tree is trained to correct the residuals of the previous ones. This method is particularly effective in complex datasets as it methodically minimizes errors and improves prediction accuracy over iterations. In contrast, Random Forest involves training multiple decision trees independently and averaging their predictions, which can be less effective in datasets where sequential error correction is beneficial.

The enhanced performance of XGradient Boosting in handling complex patterns and interactions within our dataset justified its selection over other methods such as Random Forest, despite the latter's robustness and popularity in various applications.

### 3.1. Enhanced gradient boosting with XGBoost

The XGBoost algorithm, used in this study, is a refined version of gradient boosting regression Developed by Chen and Guestrin (2016), known for its high efficiency and robust regularization techniques. The methodology involving XGBoost encompasses the following aspects:

- **Objective Function:** XGBoost optimizes a comprehensive objective function that integrates both the loss function and regularization to mitigate overfitting and enhance model performance. The objective function is meticulously crafted as follows:

$$\mathcal{L}(\phi) = \sum_i l(\hat{y}_i, y_i) + \sum_k \Omega(f_k) \quad (1)$$

where $\mathcal{L}(\phi)$ represents the total loss to be minimized. The first term, $\sum_i l(\hat{y}_i, y_i)$, quantifies the prediction error through a loss function $l$, contrasting the predicted outcomes $\hat{y}_i$ against the actual targets $y_i$. The second term, $\sum_k \Omega(f_k)$, introduces regularization, encapsulating the complexity of the model to prevent overfitting. This dual approach ensures a balanced focus on achieving high predictive accuracy while maintaining model simplicity.

- **Regularization Component:** The regularization term $\Omega(f_k)$ plays a pivotal role in XGBoost's formulation, explicitly designed to control model complexity:

$$\Omega(f_k) = \gamma T + \frac{1}{2}\lambda\|w\|^2 \tag{2}$$

  Here, $T$ signifies the total number of leaves in the tree $k$, and $w$ denotes the vector of scores on the leaves. The parameter $\gamma$ serves as the complexity cost assigned to each additional leaf, penalizing the growth of the tree structure. The term $\lambda$ is the L2 regularization parameter on the leaf weights, reducing their magnitude to enhance generalization capabilities. Collectively, these components of $\Omega(f_k)$ ensure that the model remains both effective and interpretable.
- **Gradient and Hessian Optimization:** The optimization strategy in XGBoost employs the Newton–Raphson method, leveraging second-order derivatives for a refined adjustment of the model. The approximation for optimization at iteration $t$ is given by:

$$\tilde{\mathcal{L}}(t) = \sum_{i=1}^{n}\left[g_i f_t(x_i) + \frac{1}{2}h_i f_t^2(x_i)\right] + \Omega(f_t) \tag{3}$$

  In this equation, $g_i$ and $h_i$ represent the first and second-order gradients of the loss function with respect to the predictions, facilitating a nuanced understanding of the error landscape. This methodological choice allows XGBoost to more accurately pinpoint the direction and magnitude of model updates, markedly improving convergence speed and prediction accuracy.
- **Shrinkage and Column Subsampling:** Beyond regularization, XGBoost incorporates shrinkage and column subsampling to further enhance model robustness. Shrinkage, through the application of a learning rate, systematically scales down updates to the model, enabling more gradual learning and reducing the risk of overfitting. Column subsampling introduces randomness into the feature selection process for tree construction, promoting diversity among the trees and contributing to a more generalized model performance.

The implementation of XGBoost in our study uses these components to create a predictive model that is accurate and capable of handling the complexities inherent in geothermal gradient data.

### 3.2. Model building and evaluation

The development and validation phases of the model were structured to enhance its learning efficacy and accuracy in predicting geothermal gradients. Our methodological framework incorporates several critical steps that collectively improve the model's performance.

The initial phase involved partitioning the data set into training and testing subsets with an 80/20 split, a standardized step for model construction, and subsequent validation with independent test data. This split facilitates the proper evaluation of the model's ability to generalize. Furthermore, we performed hyperparameter optimization to fine-tune the XGBoost model, enhancing its predictive accuracy.

Following the hyperparameter tuning, the XGBoost model was trained using a squared error loss function. The training process incorporated a five-fold grid search approach to fine-tune parameters such as shrinkage, maximum tree depth, and subsampling rate. This iterative refinement ensured the robustness and adaptability of the model.

The validation of the model used a quantitative framework influenced by Lösing and Ebbing (2021) and Rezvanbehbahani et al. (2017), focusing on absolute and relative error metrics. For the assessment of the absolute error in the test set, we computed the mean square error (RMSE) and mean absolute error (MAE). Furthermore, normalized root mean square error (nRMSE) and normalized mean absolute error (nMAE) were calculated to provide standard and relative error measurements that are invariant to rescaling, facilitating comparative analysis.

The coefficient of determination, or R-squared ($R^2$), was calculated for both the test and training sets. The $R^2$ for the test set provides information on how well the predicted geothermal gradients align with the actual observed values, indicating the accuracy of the model in unseen data. Similarly, the $R^2$ value for the training set, referred to as the model score, reflects the proportion of variance in the geothermal gradient data that is predictable from the input variables, providing a gauge of the model's efficacy during the learning phase.

Additionally, the process included analyzing the feature importance for each tree, averaged across the model. This analysis revealed how each feature reduced the uncertainty at its split point, weighted by the observation count in each node. These insights were crucial for improving the prediction model and understanding the factors that affect the predictions of geothermal gradients.

## 4. Data and features

In this section, we explain the data used to build our predictive model. The target feature $y$ was the Apparent Geothermal Gradient (AGG) measurements from over 4000 well logs along the colombian territory, and for the predictor features $X$ we use certain geological and geophysical features which are linked to each AGG value.

### 4.1. Apparent geothermal gradient

Colombian AGG measurements were taken from the Alfaro et al. (2009) Colombian Geothermal Gradients map, which includes the data from over 4000 wells (Fig. 5(a)). The methodology they adopted for calculating the geothermal gradient for each well involved the following steps:

1. **Measurement of Bottom Hole Temperature (BHT):** BHT measurements from gas and oil wells were primarily used. These measurements represent the maximum temperature recorded in a well. Due to the drilling process, these measurements are often lower than the actual formation temperature.
2. **Temperature Measurement Correction:** Various models and methods were used to correct for the effects of drilling on formation temperatures. The most widely used method for correcting BHT data was the AAPG method, which compares BHT measurements with equilibrium temperatures in wells in Louisiana and West Texas. The correction was achieved using a third-order polynomial equation, with BHT correction as a function of depth.
3. **Data Sources and Treatment:** Data were obtained from the Finder and LogDB tables supplied by EPIS. The organization of data involved forming pairs of depth-temperature data, and the selection process involved criteria for data quality. Units were converted from degrees Fahrenheit and feet to degrees Celsius and kilometers. Temperatures were corrected using the empirical method of AAPG.
4. **Estimation of Individual Well Gradients:** The maximum gradient within the upper crust was assumed to be vertical and calculated using the equation $G = \Delta T/\Delta P = (TBHT - Ts)/Z$, where $TBHT$ is the corrected bottom hole temperature, $Ts$ is the surface temperature, and $Z$ is the depth. The surface temperature was estimated on the basis of correlation with elevation.
5. **Interpolation and Map Creation:** A minimum curvature interpolation method was used with specific parameters in the Geosoft software. The map was edited in ArcMap, including official coverages of sedimentary basins and coverage of points with conventions for differentiating the magnitude of the estimated gradient and the measured BHT depth.

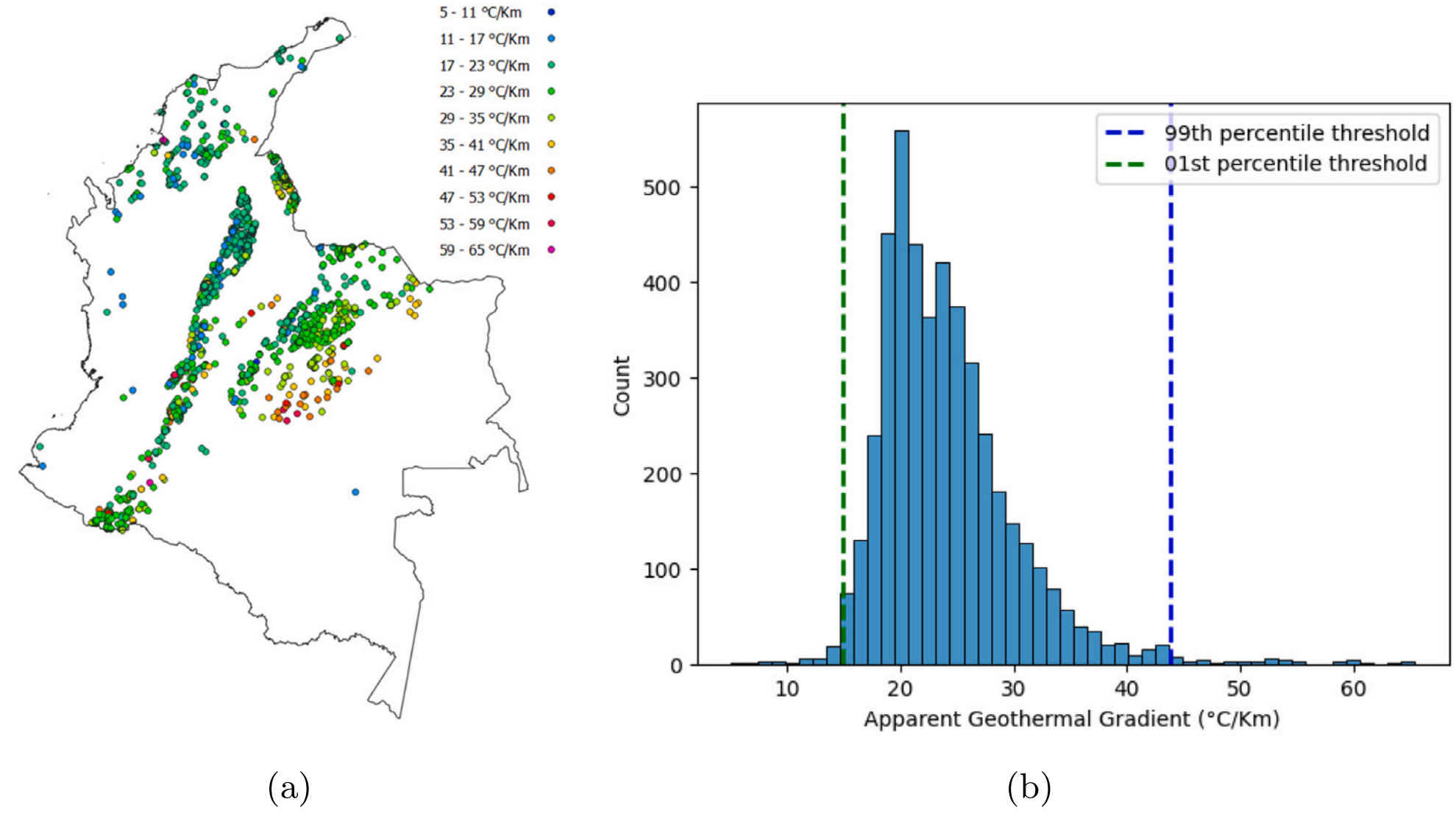

**Fig. 5.** (a) Measured geothermal gradients in Colombia, (b) Histogram of apparent geothermal gradient values indicating a right-skewed distribution. Dashed lines indicated the representative percentiles for lowest and extreme gradients that will be discarded.

Note that the gradient values obtained by Alfaro et al. (2009) represent the average gradient over certain depth measured for each well. These apparent gradients are important for indicating broader regional trends and identifying areas with geothermal anomalies (Fig. 4a). However, they may not precisely reflect localized thermal behaviors at specific depths. This is an important consideration for interpreting gradient data and applying it in our predictive model.

The geothermal gradient values are predominantly ranging between 10 and 40 °C/Km, as shown in Fig. 5(b). The data exhibit a unimodal distribution, characterized by a pronounced peak and a right-skewed asymmetry, indicating a longer tail of higher geothermal gradient values.

The distribution suggests that we are working with an imbalanced dataset, while the majority of measurements are clustered at lower gradients, a subset of regions exhibits elevated gradients, potentially highlighting areas of heightened geothermal activity. This skewness towards higher values contributes to an increased error rate in the model when predicting in these less represented higher gradient zones.

To deal with underrepresented high-gradient values, we defined a frequency-based weighting so that the less-frequent gradients within a 15-bin histogram distribution have the higher weights, and the most-frequent gradients have the lowest weight (Fig. 6). The number of bins and, therefore, the values of the weights; were adjusted based on model evaluation.

Furthermore, it is important to consider that the geothermal gradient data have an inherent bias: Although the measurements were made mainly in basins, a considerable amount of the data is close to Andean and volcanic areas, so measurements made in areas farther away tend to be underrepresented and affected the prediction for areas where there are no measurements. To correct for this bias, we defined each high-gradient well located more than 75 km from the nearest volcano as "non-volcanic" and assigned a weigh of 8 to those wells.

Finally, we constrained the training and test sets to a specific range: from the 1st percentile value of 14.87 °C/km to the 99th percentile value of 43.80 °C/km (Fig. 5(b)). This selection ensures that the model is trained and evaluated in a confidence range.

### 4.2. Geological and geophysical data

To predict the apparent geothermal gradient, we utilized a diverse array of geological and geophysical data. This data provided insight into the subsurface conditions necessary for accurate geothermal gradient estimation.

The AGG dataset includes latitude and longitude, for georeferencing well sites. This geospatial positioning is used to integrate the spatial variations of the geothermal data into the model. Alongside these coordinates, topography and elevation play a significant role. The topographic elevation data, derived from well data and NASA's global Earth Digital Elevation Model (DEM) as of 2021, provide a vertical reference point which is essential for geothermal gradient measurements.

Additionally, subsurface characteristics are another key aspect of the dataset. Insights into the thermal structure of the lithosphere and the behavior of heat at depth are garnered from the depths to the Moho discontinuity and the Curie isotherm. These are sourced from studies by Quintero et al. (2014) and Uieda and Barbosa (2017). Moho discontinuity and elevation data are strictly correlated to litosphere thickness, which is a great predictor for the goeothermal gradient in the continent (Kolawole and Evenick, 2023); Studies from Quintero et al. (2019) and Ponce and Hernández (2014) show that there is an inverse correlation between the thermal gradient and the Curie isotherm. The depth of lithosphere-asthenosphere boundary was regarded as one of the top five important features in previous models (e.g., Rezvanbehbahani et al., 2017; Lösing and Ebbing, 2021), however, it was discarded because of the lack of a representative dataset for Colombia.

Additionally, the presence of geophysical anomalies significantly influences the dataset. Magnetic anomalies, documented in the WDMAM V2 map by Dyment et al. (2020), are important as they potentially correlate with AGG. The deepest magnetic sources can be associated to the Curie depth (Lösing and Ebbing, 2021).

Gravity anomalies are also strong indicators of crustal composition. We use Free Air anomalies and Vertical Gravity Gradients from Sandwell et al. (2021) and Pavlis et al. (2012) datasets, it can indicate variations in the subsurface composition and structure. These variations are essential in understanding the distribution of subsurface heat and are influenced by crustal properties, such as the density variations and their correlation with rock types. The Bouguer anomaly was calculated using the Free Air anomaly with the Bouguer correction, and using this parameter allows us to reduce the elevation influence in the gravitational anomaly. After testing, we decided to keep the three gravity features.

The dataset also considers fault data, specifically the proximity to faults, calculated using data from Gómez-Tapias et al. (2020) and analyzed through a Nearest Neighbors Algorithm. This proximity provides insights in evaluating potential permeability and fluid flow, which can significantly affect subsurface thermal regimes. Active faults, in particular, identified by Veloza et al. (2012), are of immense importance due to their potential role in localizing geothermal systems.

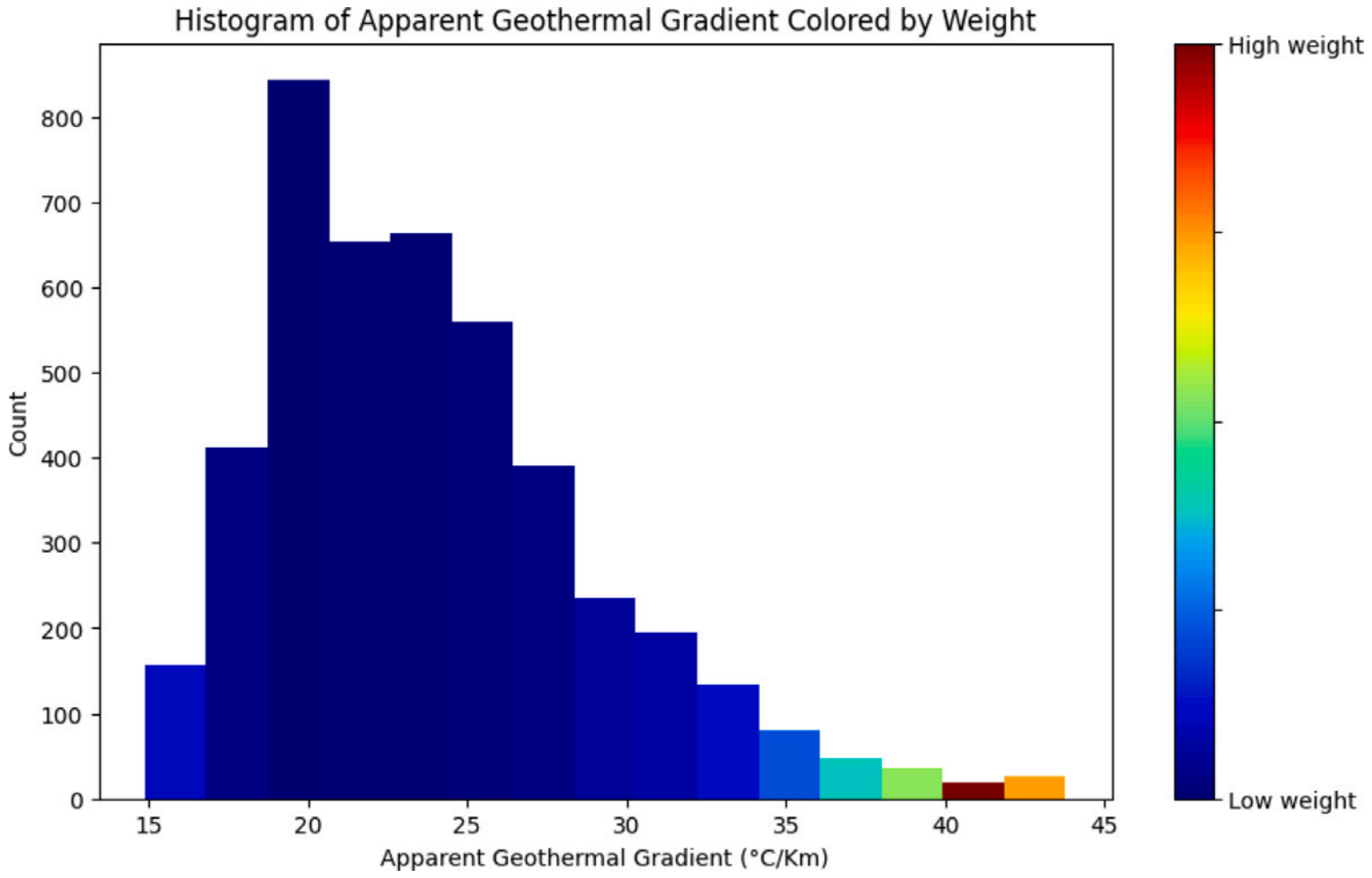

**Fig. 6.** 15-bin frequency histogram showing the weighting distribution for the geothermal gradients. The weights were normalized so the highest weight is 1.0 and the lowest is 0.02.

**Table 1**
Geological and geophysical features used in the study with their respective sources.

| Feature | Description | Source |
|---|---|---|
| Elevation (m) | Height above sea level. | NASA Visible Earth (2021) |
| Moho Depth (m) | Depth to the Mohorovičić discontinuity. | Uieda and Barbosa (2017) |
| Curie Depth (Km) | Depth at which rocks in a specific geographical area encounter the Curie temperature | Quintero et al. (2014) |
| Magnetic Anomaly (nT) | Earth's magnetic field anomaly at the point. | Dyment et al. (2020) |
| Free Air Anomaly (mGal) | Gravity anomaly after Free Air correction | Sandwell et al. (2021) |
| Bouguer Anomaly (mGal) | Gravity anomaly after Free Air correction and Bouguer corrections | Sandwell et al. (2021) |
| Vertical Gravity Gradient (E) | rate of change of vertical gravity ($gz$) with height ($z$) | Pavlis et al. (2012) |
| Faults (m) | Distance to the nearest normal, reverse or strike-slip fault. | Gómez-Tapias et al. (2020) |
| Active Faults (m) | Distance to the nearest active fault. | Veloza et al. (2012) |
| Distance to basement (m) | Surface distance to igneous or metamorphic outcrops | Gómez et al. (2015) |
| Volcanic Domain | Boolean variable to define whether the well is located in a volcanic zone or not. | Gómez-Tapias et al. (2020) |
| AGG measurements (°C/Km) | Target variable, measured from bottom hole data | Alfaro et al. (2009) |

Lastly, the depth to the basement rocks (igneous and metamorphic rocks) is a critical factor in understanding the influence of subsurface rock types on thermal behavior, as they are a main source of radiogenic heat production compared to sedimentary rocks (Hasterok et al., 2018). Due to the absence of a specific basement depth model, an alternative method is employed using the geological map by Gómez et al. (2015). This method involves classifying rock outcrops as either basement (igneous and metamorphic) or non-basement (sedimentary and deposits), and measuring the distance from each well site to the nearest basement outcrop. This assessment, along with other features such as faulting or structural constraints that may affect the heat flow, helps to understand the impact of these rocks on geothermal gradients.

Integration of geospatial coordinates, topographical data, subsurface characteristics, geophysical anomalies, fault data, and proximity to basement rocks forms a comprehensive framework for accurately modeling and understanding the subsurface geothermal behavior, reflecting the multidisciplinary nature of geothermal research. Each of these factors contributes uniquely to understand the gradient distribution, underlining the complex interplay of geological, geophysical, and spatial factors in geothermal exploration and modeling.

Detailed descriptions of these variables are provided in Table 1 and two graphical examples are shown in Fig. 7. The correct interpretation and use of these data is crucial for predictive modeling of geothermal gradients across Colombia. The features were directly interpolated to obtain values at specific geographic points without resampling, the dataset with the lowest resolution (Moho Depth) has a 0.4° latitude-longitude grid.

### 4.3. Hyperparameter tuning

To optimize the performance of the model, we defined a grid of hyperparameters. The code iterates through these hyperparameter combinations, training the model, and evaluating mean squared error (MSE) on the test set for each configuration. The process systematically identifies the combination of hyperparameters that minimizes MSE, ensuring the model is fine-tuned for optimal predictive accuracy on the given dataset (Table 2).

## 5. Results

The result we present for this study is a predicted geothermal gradient map for all the Colombian territory. The accuracy of the methodology is evaluated using the test subset derived from the initial dataset partitioning. We examine the key performance metrics: MAE, RMSE, nMAE, nRMSE and the coefficient of determination ($R^2$) to quantify the predictive precision of the model. Additionally, we employ visualizations to facilitate a comprehensive analysis: plots for actual versus predicted values, diagrams illustrating feature importances, and assessments delineating the influence of individual features on the model's output.

A geospatial comparison of actual data against model predictions is conducted to discern the model's spatial prediction accuracy. This comparison not only substantiates the model's validity, but also underscores any region-specific deviations.

### 5.1. Model validation

The initial phase of the model evaluation involved scrutinizing its performance in both the training and the test subsets, the latter representing previously unseen data. For the training set, the model achieved an MAE of 1.4793, an RMSE of 2.1371, and a coefficient of determination ($R^2$) of 0.8287. These metrics are typical for models evaluated on their training data. Conversely, the test data got an MAE of 2.6826, an RMSE of 3.5842, and an $R^2$ of 0.5497.

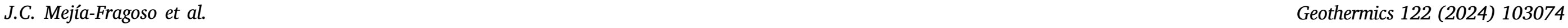

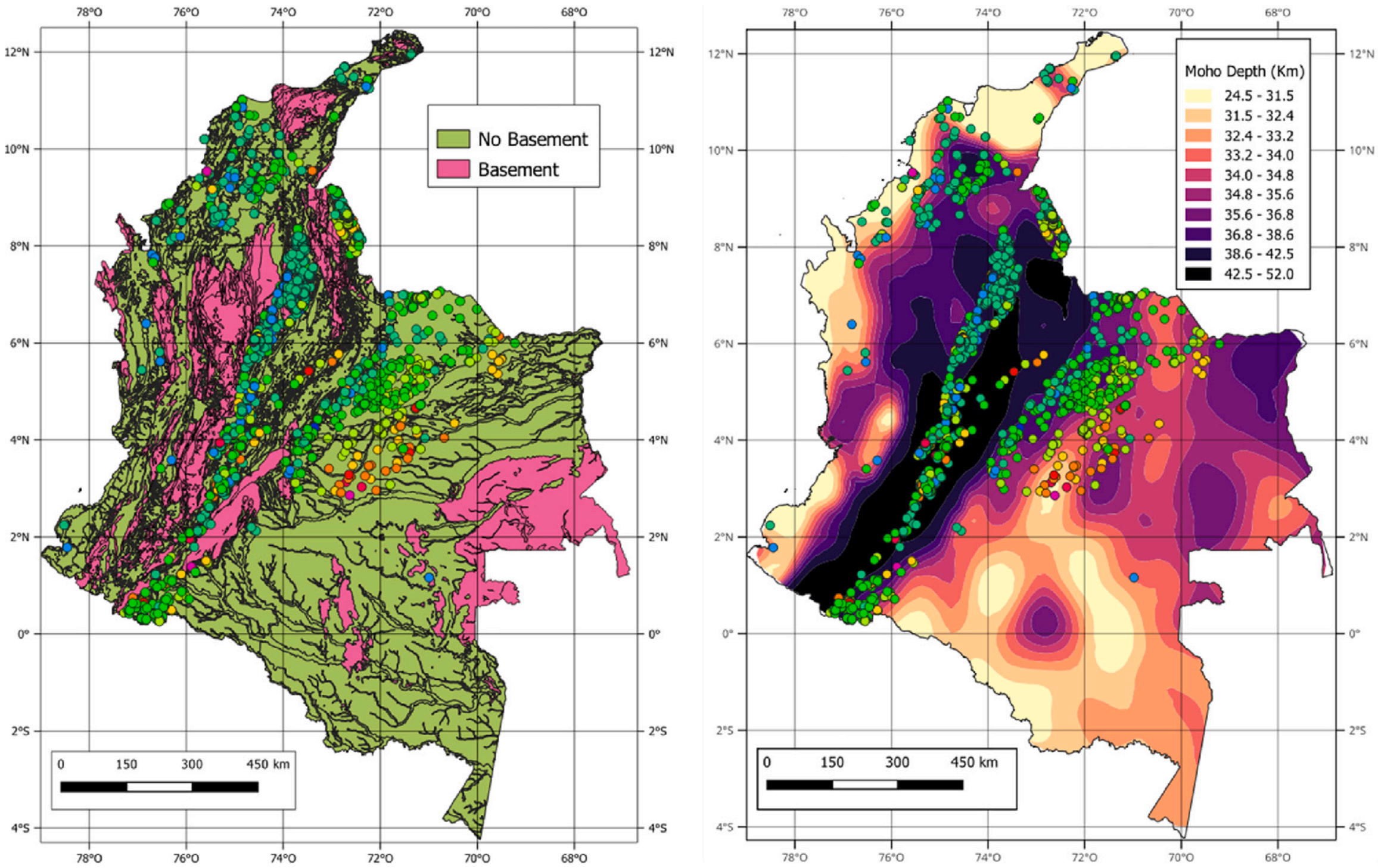

**Fig. 7.** Side-by-side maps of Basement rocks outcrops and Moho Depth with the measured gradient locations.

**Table 2**
Best Hyperparameters after Model Tuning.

| Hyperparameter | Value | Meaning |
|---|---|---|
| `max_depth` | 10 | Maximum depth of a tree. |
| `learning_rate` | 0.1 | Step size shrinkage used in updates. |
| `gamma` | 0.0 | Minimum loss reduction required to make a further partition. |
| `min_child_weight` | 1 | Minimum sum of instance weight needed in a child. |
| `colsample_bytree` | 0.5 | Fraction of features to be randomly sampled for each tree. |
| `reg_alpha` | 0.5 | L1 regularization term. |
| `reg_lambda` | 0.5 | L2 regularization term. |
| `subsample` | 1 | Fraction of training data to be randomly sampled for each boosting round. |

The disparity in $R^2$ could be attributed to several factors, including the underrepresentation of higher geothermal gradient values, the geographic bias of the measurements and the inherently complex nature of the dataset. The test set's normalized error metrics, nMAE and nRMSE, were recorded at 0.06 and 0.12 respectively, indicating that our predictions are expected to have a deviation from 6% to 12% of the correct values.

A detailed examination of the residuals, as depicted in Fig. 8, reveals a random distribution around the zero line. This pattern suggests an absence of bias in the residuals, corroborated by a residual mean close to zero, indicating that the model is on average neither overestimating nor underestimating significantly. However, points with high residuals highlight instances where model performance was suboptimal, though most residuals cluster near zero.

Regarding the plot of actual versus predicted values, illustrated in Fig. 9, the concentration of points along the ideal prediction line indicates proficient modeling of the target values. The dispersion of points as the target value increases suggests increasing variance in the model's predictions. The model demonstrates higher accuracy for lower actual values while tending to underestimate higher actual values in the test set. The overlap of training (blue) and test (red) data points

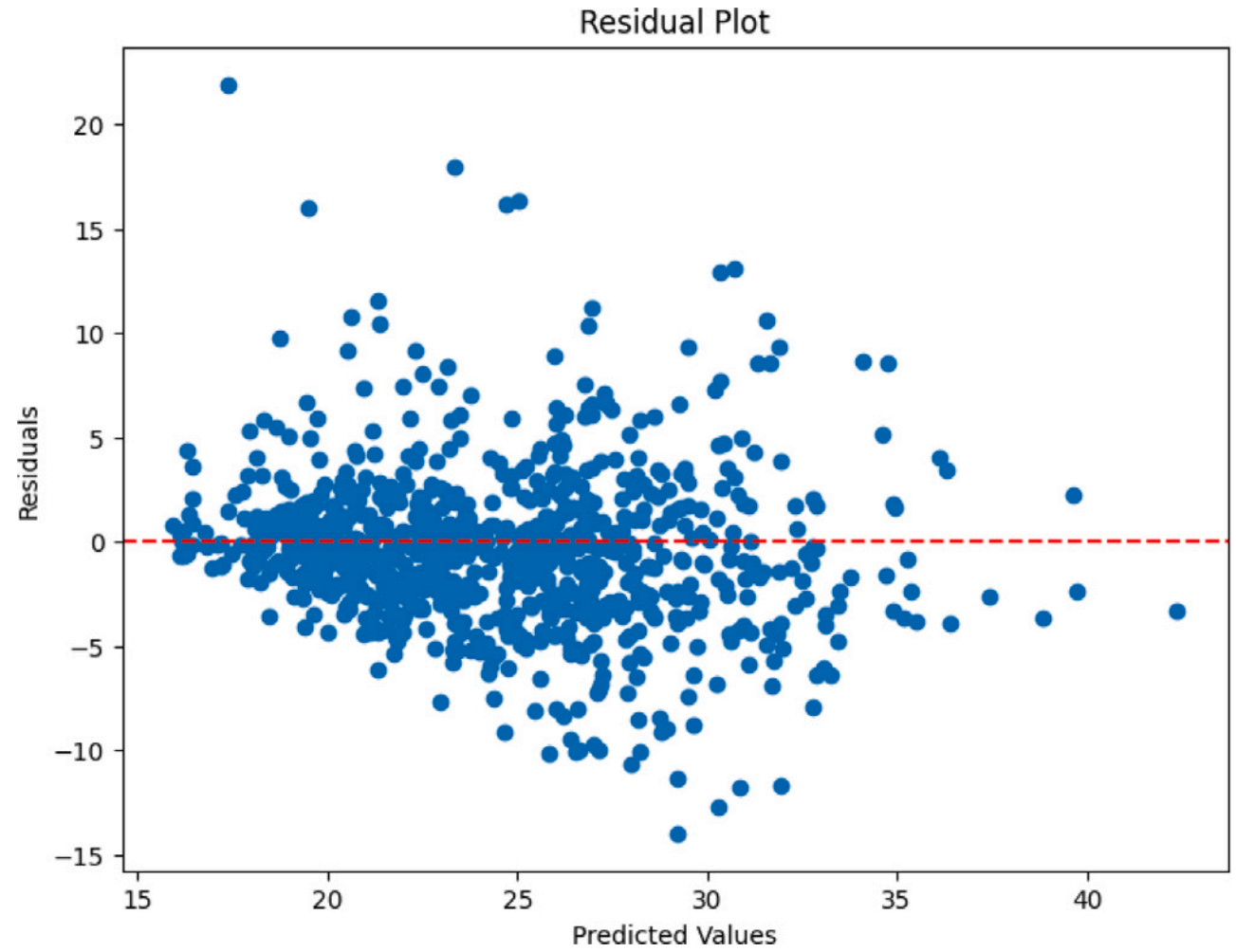

**Fig. 8.** Residuals distribution of the model.

in the plot is indicative of good model generalization, evidenced by the similarity in performance across the training and test datasets. This overlap is a positive sign, suggesting that the model is not overfitting

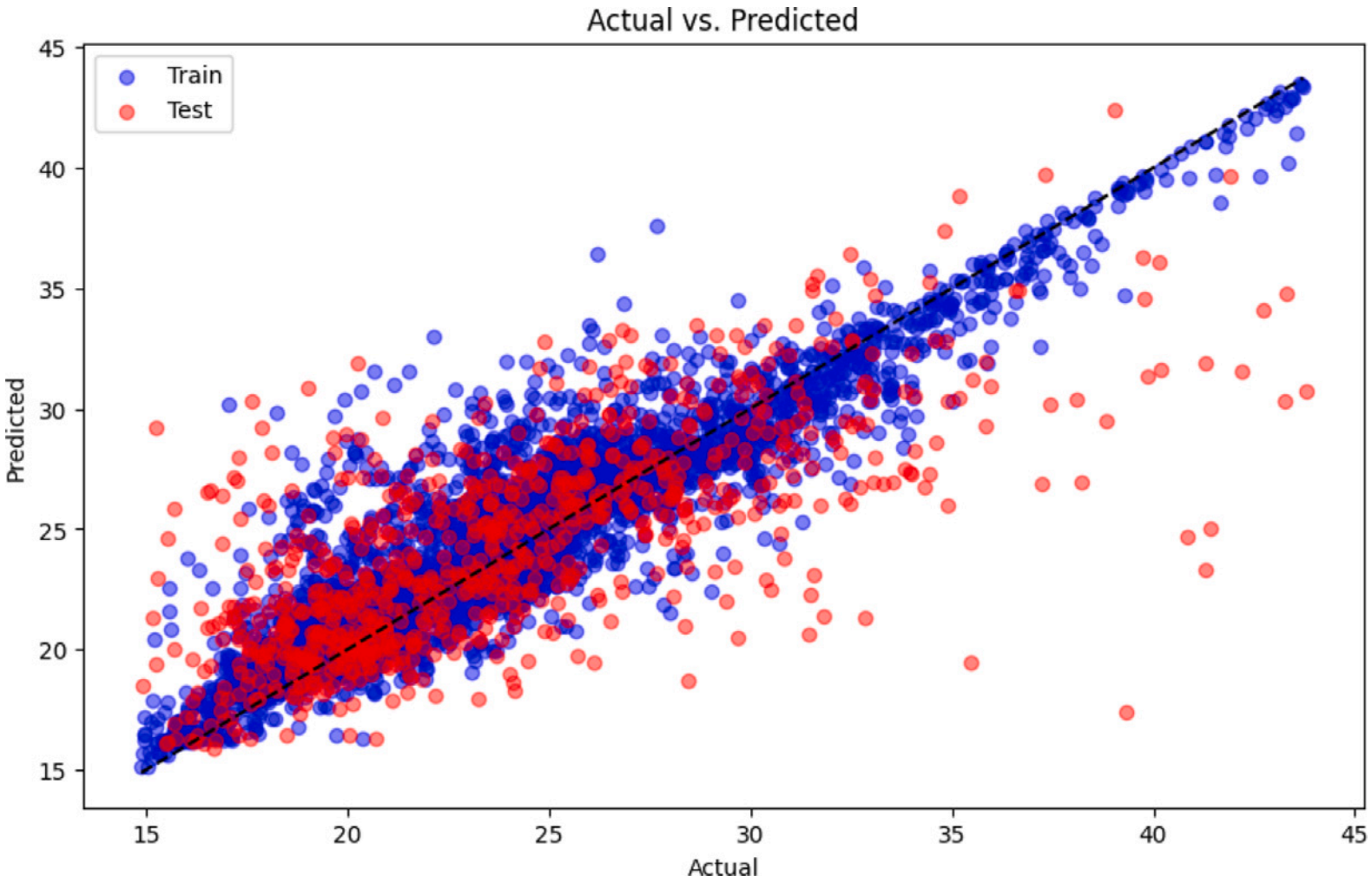

**Fig. 9.** Actual vs. Predicted values plot comparing the training and test subsets.

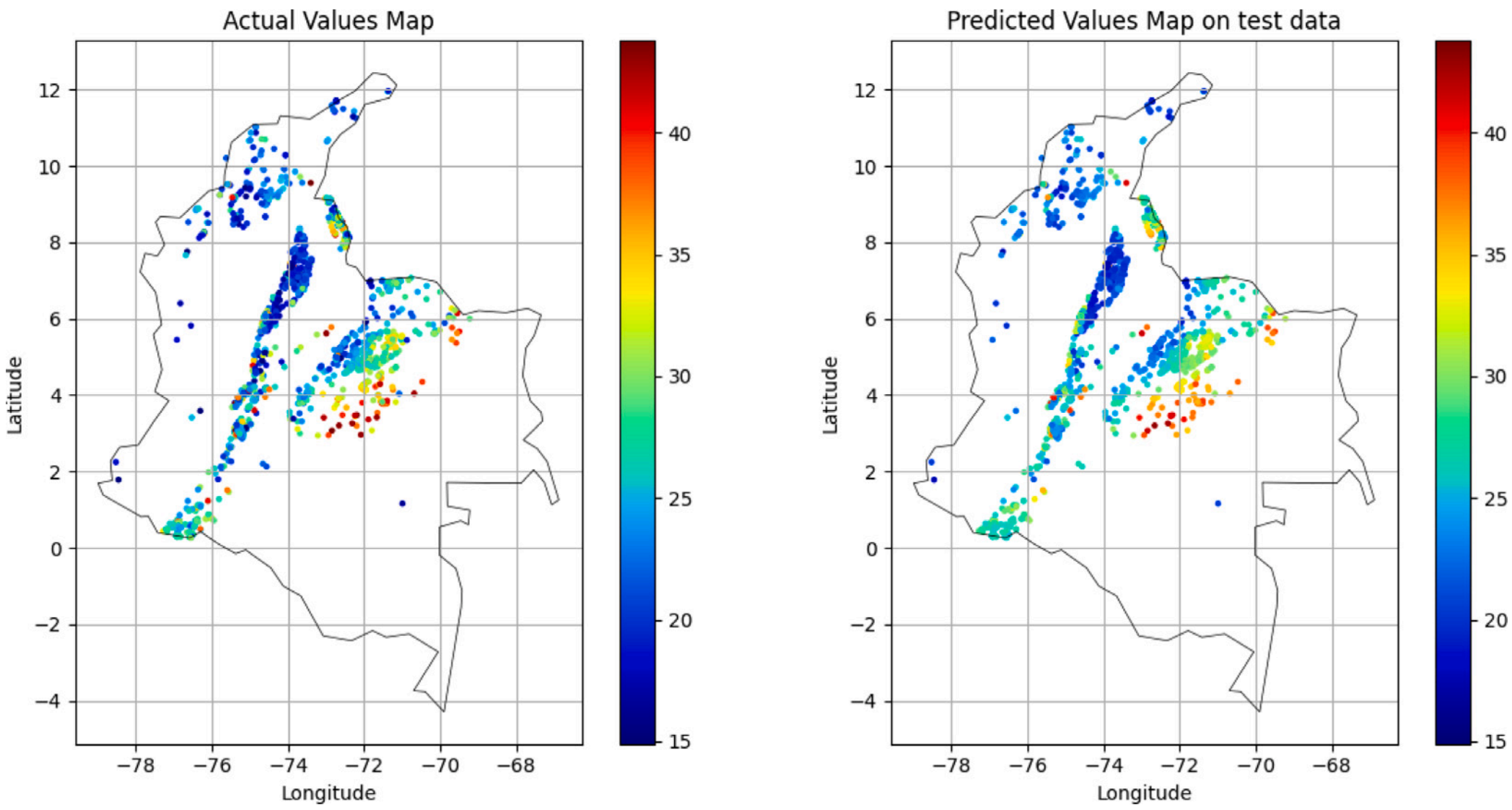

**Fig. 10.** Side-by-side comparison of actual and model-predicted geothermal gradient values, highlighting the model's spatial prediction accuracy.

to the training data and most of the variance in the test set is towards the high gradient values.

Evaluating the model's geospatial predictive performance is crucial for understanding its applicability in practical scenarios. This involves a comparative analysis of observed geothermal gradient values against the model's predictions.

Additionally, a side-by-side visualization of the actual and predicted geothermal gradient values was carried out, as shown in Fig. 10. The actual values map is a representation of empirically measured geothermal gradients used for the training, while the predicted values map is derived from the model estimations. The consistency of geothermal gradients across both maps serves as an indicator of the model's spatial accuracy.

The Fig. 11 highlights the differences between the actual and the predicted geothermal gradient values, showing regions where the model's predictions deviate from measured data. Those discrepancies suggest potential areas for model enhancement or may pinpoint inherent limitations within the data or the model's structural assumptions. The tendency of the model to underestimate geothermal gradients, specially when they are higher, indicates possible gaps in its representation of heat source distribution or heat transfer mechanisms, and may require data enhancement.

Comparing actual with predicted values yields critical insights into the model's predictive strengths and weaknesses. Such an assessment is imperative to refine the model's predictive capabilities, thereby enhancing its contribution to geothermal exploration and resource evaluation.

### 5.2. Feature importance analysis

Feature importance provide insight into which variables are most informative for making accurate predictions. Multiple techniques were used to determine the importance of the features, each offering a unique perspective on the data.

*Weight-based importance.* The weight-based feature importance, as shown in Fig. 12, indicates that active faults, elevation and Moho depth are the most frequently used features in the model to make decisions, with structural features and basement proximity having also a high importance. Potential fields and Curie Depth, on the other hand, are less frequently used. The importance is equilibrated among most of the features. This highlights the model's reliance on all the data to predict the target variable, especially topography, subsurface discontinuities, and structural influence.

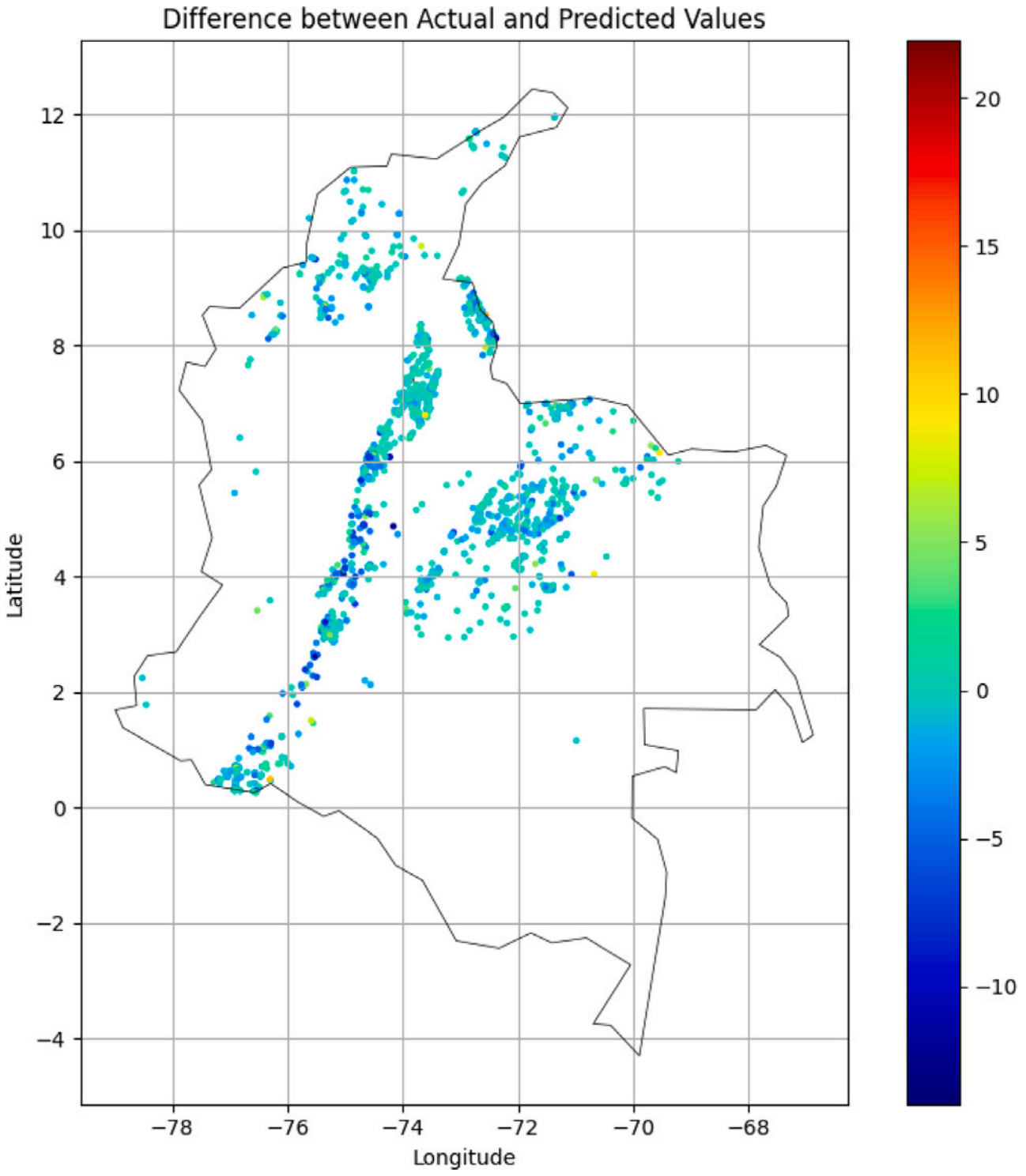

**Fig. 11.** Differences between actual and model-predicted geothermal gradient values.

*Gain-based importance.* Fig. 13 presents the gain-based feature importance, which reflects the improvement in accuracy brought by a feature to the branches, how the features contributed to minimize the loss function, and, subsequently, to the model's performance. Basement proximity in surface emerges as the most significant predictor, and features such as Potential fields and Curie Depth play a significant role here as well; however, most of the features importances are equilibrated.

*Mutual information.* Mutual information scores, as shown in Fig. 14, quantify the dependency between each feature and the target variable. Moho depth is seen to be highly informative, along with the geophysical and structural data.

We also evaluated each individual feature using the model score, the R2 score, the normalized root mean square error (nRMSE), and the normalized mean absolute error (nMAE). As shown in Fig. 15, each metric is visualized for every feature after training the model with each individually.

Moho depth and geophysical measurements, such as the Free Air Anomaly, consistently demonstrated high importance as isolated predictors. In contrast, the topography and Curie depth showed lower individual relevance.

Assessing the model's performance as a function of the feature count, starting with the less important (gain-based) and so on, reveals an increase in both the model score for the training set and the $R^2$ score for the test set, as shown in Fig. 16. These increments signify enhanced precision and variance explanation. Nonetheless, the progressive improvement in scores flatten after incorporating roughly nine features, indicating that further features may not substantially enhance the model's error metrics. It is essential to note, however, that each feature significantly enhances the model's spatial precision, as suggested by the feature importance and mutual information data. Thus, excluding any feature could negatively affect the outcomes: while performance levels off after nine features, the remaining features should not necessarily be dismissed. Incorporation of particularly insightful features, like the actual depth to the basement, depth of lithosphere-asthenosphere boundary or improved feature resolution could, indeed, offer advancements to the model.

The analyses suggest that a subset of features provides the most predictive power without introducing unnecessary complexity or risking overfitting. This balance is crucial for developing a model that is both accurate and generalizable.

### 5.3. Spatial distribution of predicted geothermal gradients

The spatial distribution of the predicted geothermal gradients is a key factor in evaluating the utility of the model for geothermal exploration. Fig. 17 illustrates the geothermal gradients predicted in the study area. The color gradient from blue to red indicates increasing values, and warmer colors represent a higher geothermal potential.

The map offers practical information on areas with significant geothermal activity, marking them as prime candidates for detailed exploration and development; however, it is important to understand the model limitations and, therefore, the fact that some high-gradient zones tend to be underrepresented. The effectiveness of the predictive model lies in its ability to identify these spatial variations and key trends, this characteristic making it a valuable asset in the guide of exploration strategies towards the most promising geothermal energy locations.

## 6. Discussion

The evaluation of our predictive model for the estimation of geothermal gradients has been extensive, incorporating a variety of metrics and visual analyses. Training set metrics, MAE, RMSE, and R-squared, along with the corresponding test set metrics provide a comprehensive view of the model performance. The model shows strong performance in the training set, with an MAE of 1.48, an RMSE of 2.14, and an R-squared value of 0.83, as indicated in Section 5.1. These values reflect a high degree of accuracy and a strong correlation between the predicted and actual geothermal gradients, suggesting that the model is capable of capturing the underlying patterns within the training data. However, when applied to the test set, the model's performance drops, with an MAE of 2.68, an RMSE of 3.58, and an R-squared value of 0.55, as detailed in Section 5.1. The observed increase in error and the decrease in the $R^2$ value on the test set, as opposed to the training set, may suggest possible overfitting, particularly in the context of estimating higher values of geothermal gradients where the data are less abundant. Despite this, the model's ability to generalize is demonstrated to be effective, as illustrated in Fig. 9.

Comparative analysis with related studies highlights the effectiveness of our model. For example, the study by Lösing and Ebbing (2021) in Antarctica achieved an nRMSE of 0.29 and an $R^2$ of 0.45, while Rezvanbehbahani et al. (2017) in Greenland recorded an nRMSE of 0.15 and an $R^2$ of 0.6. Our results, with an nRMSE of 0.12 and an $R^2$ of 0.55, reflect not only a lower relative error compared to these studies, but also a balanced performance in terms of the determination coefficient. This comparison highlights the accuracy and reliability of our model in different geological settings.

Our feature importance analyses, comprising weight-based, gain-based, and mutual information methods, showcased the varying impact of different features. Elevation and proximity to the basement consistently emerged as the main predictors of multiple importance measures, as discussed in Sections Section 5.2. Moho depth is also one of the most important predictors, and its relation with the lithospheric thickness suggests an agreement with the correlation of this tectonic feature with the thermal gradient defined by Kolawole and Evenick (2023). The geophysical properties, such as Bouguer or Free Air Anomaly, have been highlighted as significant by certain importance methods but appear less influential in others.

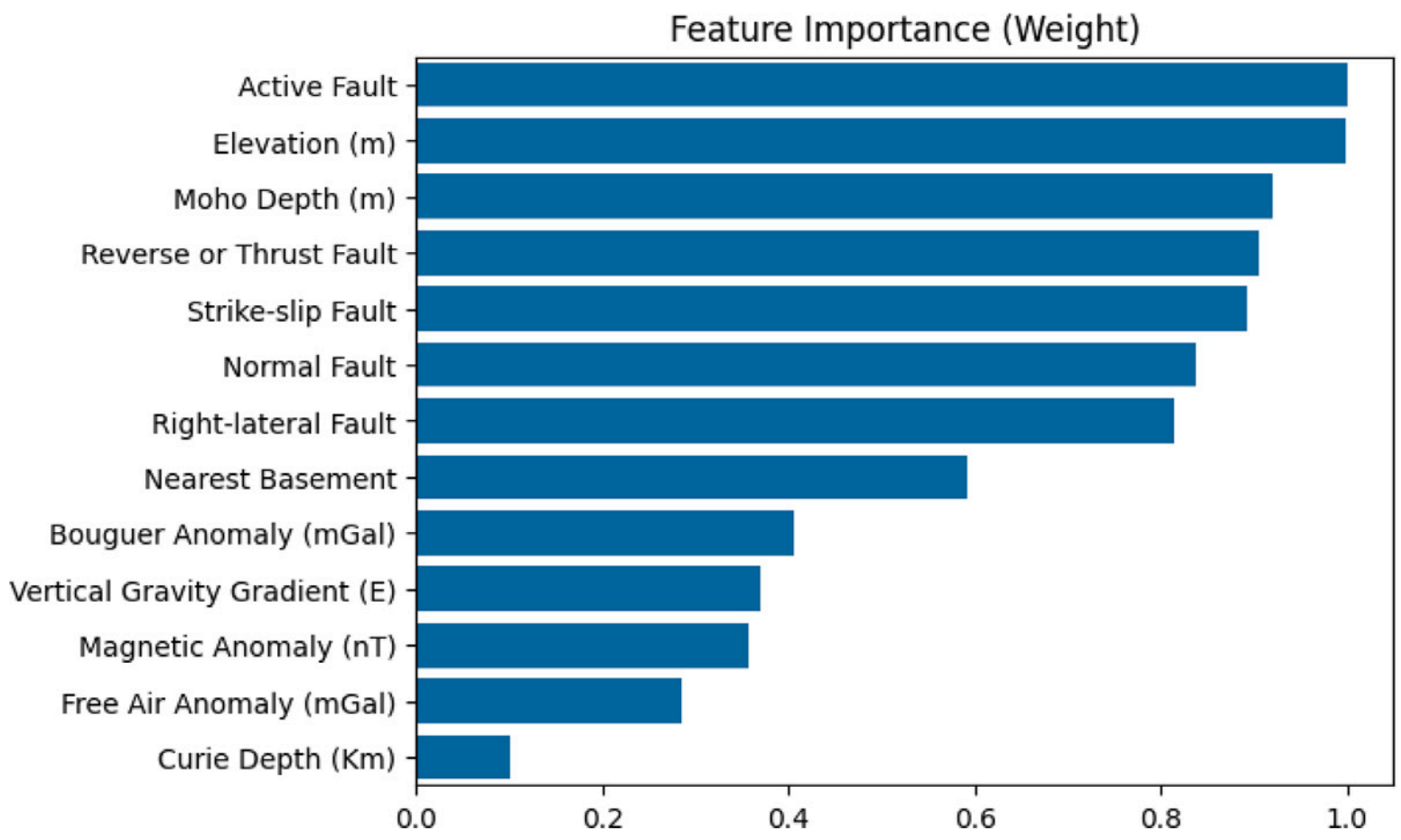

**Fig. 12.** Weight-based feature importance plot.

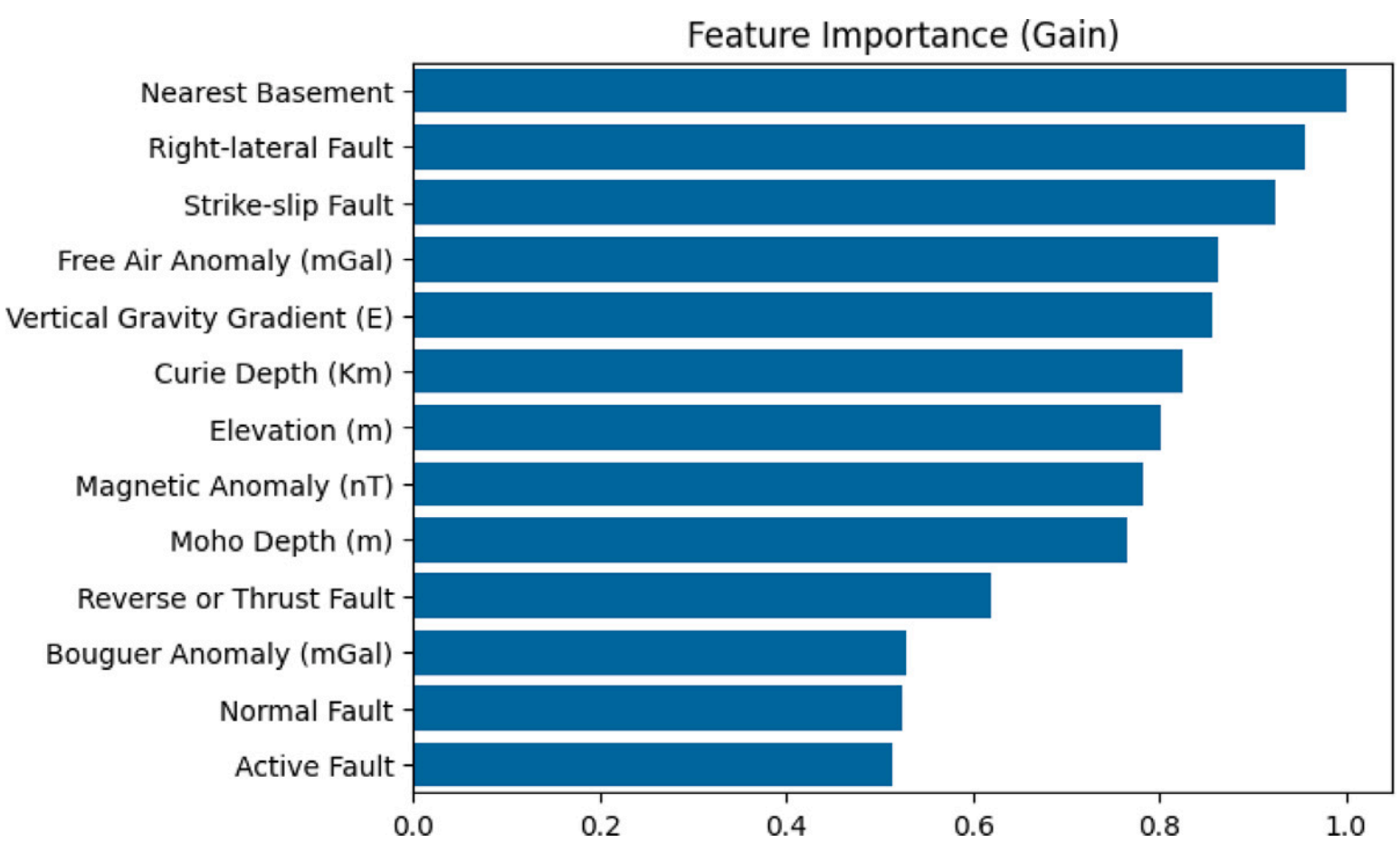

**Fig. 13.** Gain-based feature importance plot.

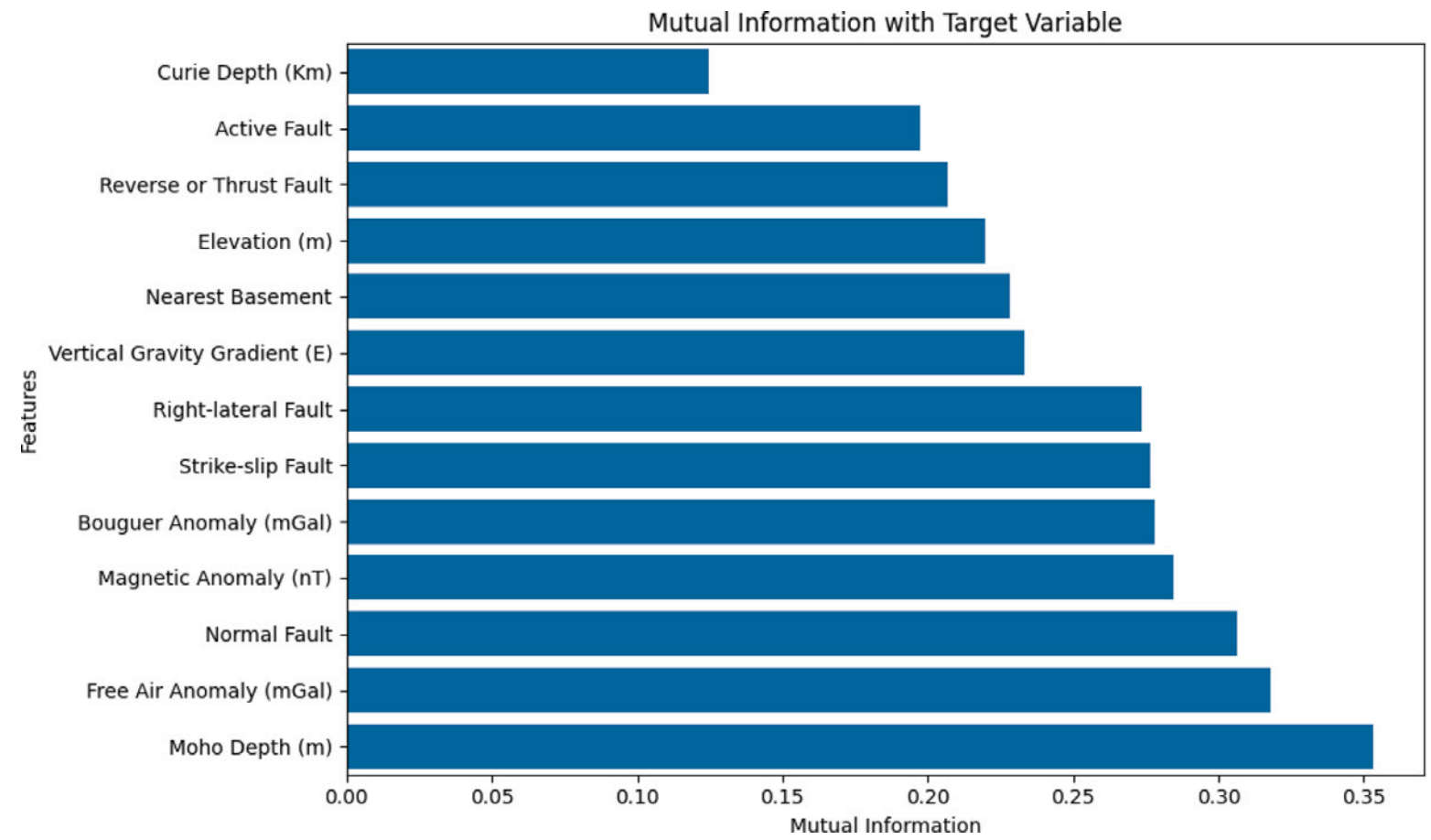

**Fig. 14.** Mutual Information with the target variable.

The observed discrepancies between training and test performance may indicate the need for further model refinement or data enhancement. Strategies such as feature engineering, parameter optimization, and integration of more specific data such as actual depth to basement or lithosphere thickness could be employed to enhance the model's predictive power and generalizability. The geospatial distribution of the predicted geothermal gradients, illustrated in Fig. 17, highlights the spatial prediction capabilities of the model. The alignment of high-gradient regions on actual versus predicted maps (Fig. 11) and the regions of significant discrepancy provide a nuanced understanding of the spatial precision of the model.

Since the geothermal gradients used for training the model are apparent gradients and were averaged to build the original dataset, we must consider that limitation and treat both the preliminary geothermal gradients map (Fig. 4a) and our predicted geothermal gradients map (Fig. 17) as indicative tools rather than definitive measures. These maps provide a pragmatic framework for estimating geothermal potential, which is particularly useful to inform future exploration studies. This approach is especially beneficial in regions where well measurements are scarce or absent, which is the target of our study.

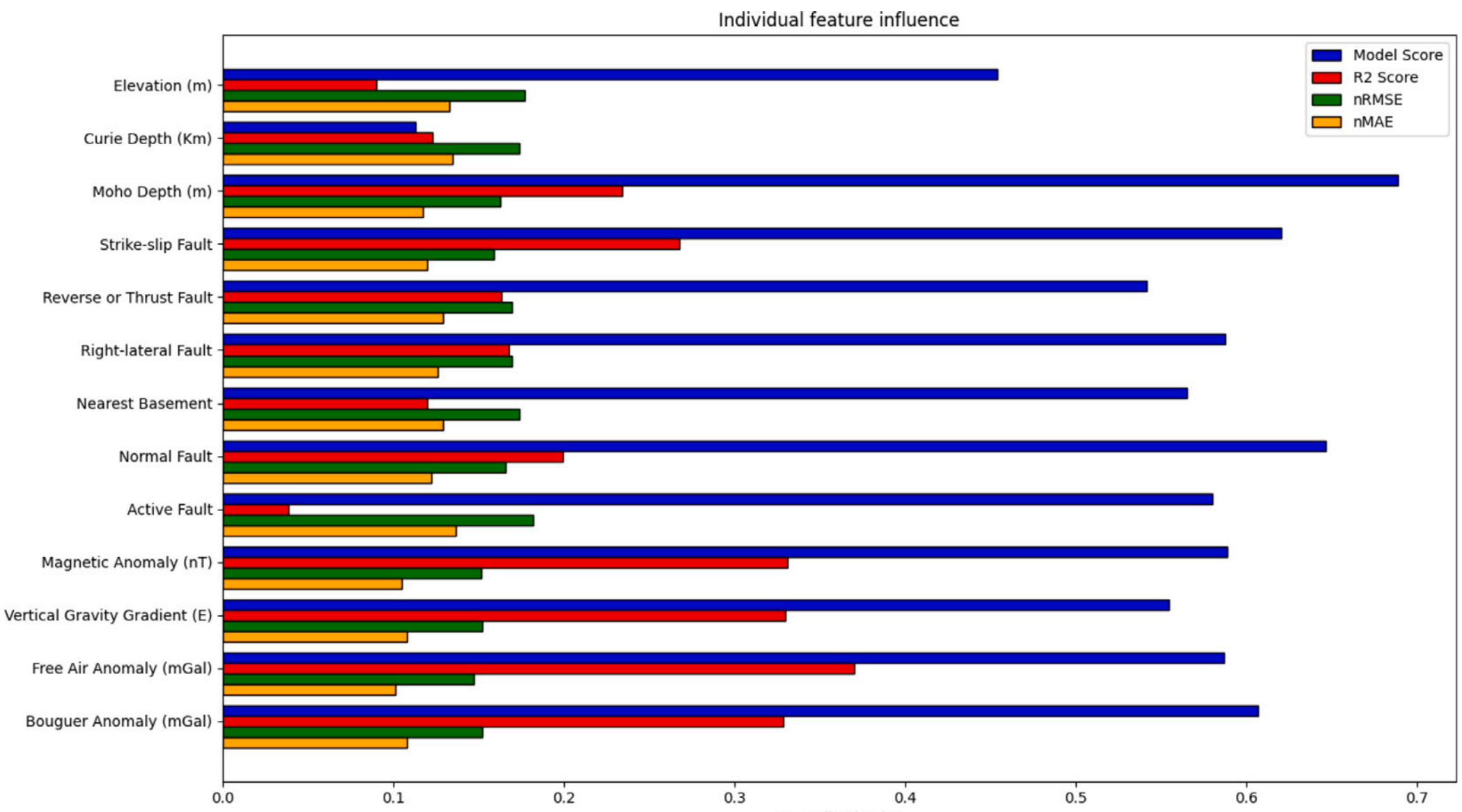

**Fig. 15.** Feature importance evaluated by various metrics for each individual feature. Values are normalized.

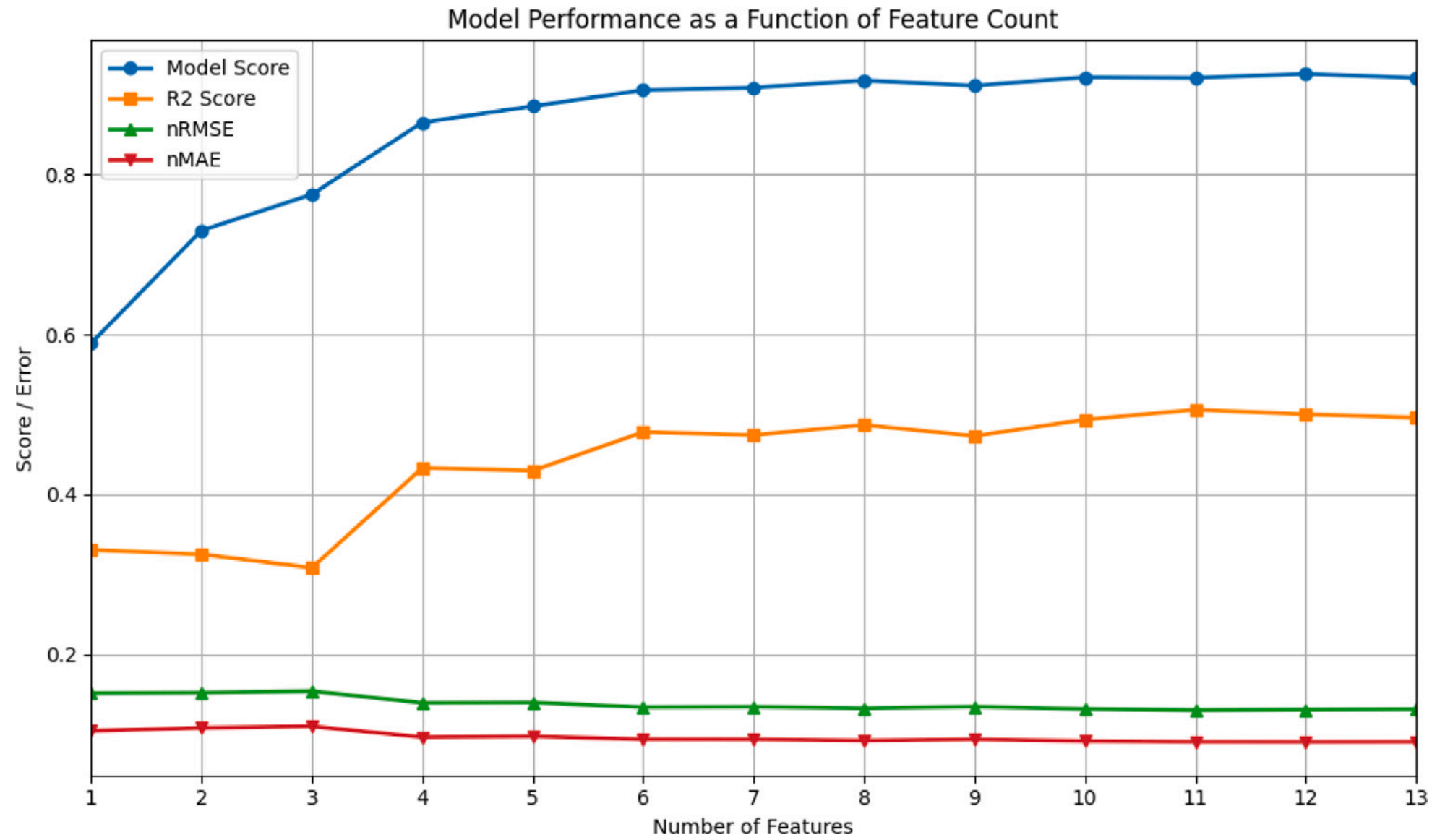

**Fig. 16.** Model performance as a function of the number of features included, starting from the less important (gain-based).

Despite the inherent limitations of the geothermal gradient data, which is depth-dependent and measured at varying depths, the model's output demonstrates notable sensitivity to geothermal indicators, and reveals high-gradient trends that strongly suggest the presence of underlying geothermal resources. These values are particularly pronounced in regions characterized by volcanic activity or where the crust is considerably thinner. The alignment of high geothermal gradients with thermal manifestations such as volcanoes and hot springs, shown in Fig. 3, and their association with fault lines detailed in Fig. 1, underscores the influence of tectonic activities on these gradients. This may indicate geothermal systems dominated by convective heat transfer, especially in areas where faulting facilitates the ascent of heat from deeper sources within the earth.

Furthermore, the geothermal gradient data display responsiveness to more stable geological zones such as the Amazon and the Guiana shield. In these regions, the crust is relatively thinner and the Moho discontinuity is shallower (Fig. 7), and the proximity of the basement to the surface enhances the efficacy of conductive heat transfer, signaling the presence of conduction-dominated geothermal systems. A standout feature of the model is its predictive accuracy in the Amazon, where data scarcity has previously been a barrier. The gradient predictions of the model for this area not only fill a critical data gap, but also exhibit remarkable agreement with the findings from the Pimentel and Hamza (2012) study of the Brazil's portion of the Amazon (Fig. 4d). This concordance is a testament to the model's robustness and its ability to extrapolate from known data to regions lacking in situ measurements.

While the predicted map tends to underestimate high gradients, as seen in previous sections, we can infer that the geothermal gradient model is effectively capturing areas with geothermal potential, especially where geological structures such as faults are prevalent, and thermal manifestations are observed (refer to Figs. 1, 2, and 3). This suggests that the geothermal model could serve as a valuable tool for targeting exploration efforts. Furthermore, the correlation between high-gradient areas and volcanic activities, as well as fault distributions and basement (igneous and metemorphic) rocks, could inform a more detailed geothermal map that includes both predicted gradients and direct geological measurements.

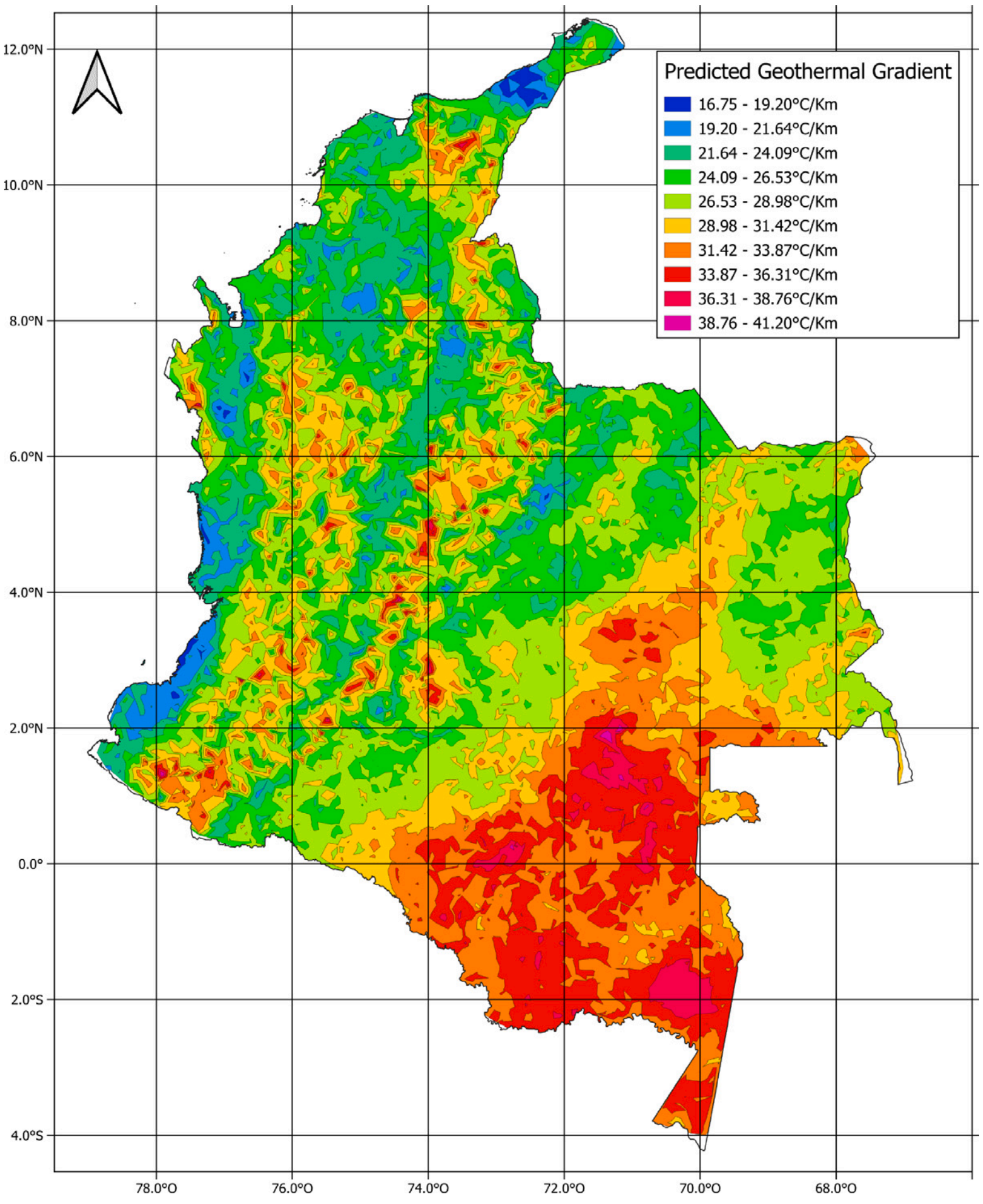

**Fig. 17.** Predicted geothermal gradient map showing the spatial distribution of geothermal gradient across the region.

Comparison between our predicted map and the preliminary geothermal gradient map (Fig. 4a) shows that, although the predicted map provides a competent representation of geothermal trends (as discussed previously), underestimating high gradients leads to significant differences in certain areas, such as the west of Eastern-Llanos and Caguán-Putumayo basins. This trend towards underestimation is likely attributed to an under-representation of higher gradient data in the training dataset. While we have attempted to mitigate this issue by assigning greater weights to these less frequent, high gradient instances (Fig. 6), it is evident that acquiring and integrating additional high-gradient data is essential for enhancing the model's accuracy in these critical areas. Further refinement in data collection and integration techniques will be crucial to improve the representation of high geothermal gradients by the predictive model.

Our predicted geothermal gradient map, when compared with the one produced by Gomez et al. (2019) for northwestern Colombia (Fig. 4b), presents some discrepancies, particularly in very-low gradient values. These discrepancies could be attributed to an under-representation of low gradient values in our dataset, as well as to the general paucity of data in the region, as illustrated in Fig. 5(a). The methodology followed by Gomez et al. (2019) might also contribute to the observed variances between their findings and the predictions of our model.

Conversely, when comparing our predicted geothermal gradient map with the one developed by Matiz-León (2023) for the Eastern-Llanos (Fig. 4c), which utilized a geo-statistical methodology, there is a notable congruence in gradient distribution. It is important to mention, however, that Matiz-León's map indicates somewhat higher gradient values. This discrepancy may arise from a variety of factors, including the different methods applied (geo-statistical modeling and interpolation over a localized area versus our machine learning approach applied countrywide) as well as under-representation of high gradient data for training our model.

It is crucial, nevertheless, to acknowledge the variations observed between the maps of Matiz-León (2023) and Gomez et al. (2019) when compared to the preliminary geothermal gradient map by Alfaro et al. (2009). Such differences not only underscore the intrinsic challenges of estimating geothermal gradients in Colombia but also how different methodological frameworks can yield slightly different results. These discrepancies could be mitigated by enriching the dataset with more direct measurements, particularly in regions that are currently data-deficient. This would enable a more refined and accurate estimation of the country's geothermal gradient landscape.

In terms of new discoveries, the model highlights areas with potentially high geothermal gradients that are not documented in existing gradient maps. These predicted hotspots, particularly in the uncharted Amazon region, present opportunities for exploration and possible development of geothermal resources. The model has proven to be a valuable asset in geothermal exploration, especially for the Amazon of Colombia, where it predicts high geothermal gradients in the absence of empirical data. Its alignment with the established geothermal gradients from Pimentel and Hamza's (2010) work, and its similarity to other gradient maps, further validates its predictive capacity.

It is also important to recognize a fundamental limitation of our model, which is inherent to machine learning models that process vast amounts of data: the challenge of accurately predicting anomalously high geothermal gradients. For example, the over 65 °C/km gradients in specific locations at sedimentary basins, or the exceptionally high gradient at the Las Nereidas well in the Nevado del Ruiz Volcano in the Central Cordillera, which approaches 140 °C/km as reported by Alfaro et al. (2009), are difficult to predict due to its rare occurrence in the dataset. Such outliers are typically treated as singular data points and may be excluded from the model's training set to maintain generalizability. Specifically, in our analysis, we focused on a confidence range between the 1st and 99th percentiles, omitting these anomalous values to avoid skewing the model with disproportionately weighted data. Therefore, measuring anomalously high geothermal gradients accurately requires more constrained, direct, and localized methodologies, tailored to specific areas where these unusual conditions may occur.

Indeed, the application of data-driven methods like Machine Learning in predicting geothermal gradients is inherently reliant on the quantity and quality of available field data. This reliance emphasizes an ongoing opportunity for enhancement, as additional data can be used to continuously refine the model. Such improvements are expected to significantly enhance the model's utility in the precise identification and assessment of geothermal resources. As the model continues to be refined, its role in the identification and assessment of geothermal resources is expected to become critical, supporting a more informed and strategic approach to energy exploration in the region.

The spatial distribution of the predicted geothermal gradients provide critical guidance for geothermal exploration. These tools enable the strategic prioritization of regions for further investigation, allowing for a focused allocation of resources to areas with the highest geothermal potential. The regions with the highest predicted gradients warrant immediate attention for exploration and data acquisition. These areas, flagged by the model, are likely candidates for sustainable geothermal development. In contrast, areas with lower predicted gradients can be investigated for baseline data collection, contributing to a more comprehensive understanding of the geothermal profile of the region.

Our predictive model can play a pivotal role in the early stages of exploration, for example, informing decision-making processes regarding site selection for geothermal power plants and drilling operations. By leveraging the model's predictions, stakeholders can enhance their exploration strategies, targeting locations with a higher likelihood of geothermal activity, and thereby advancing the exploration for renewable energy sources. This optimized approach promises not only greater efficiency in resource utilization but also a higher success rate in developing geothermal energy projects. Finally, we hope that this study and, more importantly, the open training dataset we have provided, will motivate other researchers to explore additional machine learning and data-driven techniques, which without a doubt would lead to improved models.

## 7. Conclusion

This study successfully developed and validated a machine learning model using a Gradient-Boosted Regression Tree algorithm to predict geothermal gradients across Colombia. The effort marks a significant advancement in geothermal energy exploration, particularly within areas of Colombia that have previously lacked sufficient geophysical and geological data.

By employing a diverse array of geological and geophysical data, our model has demonstrated its capacity to provide accurate predictions of geothermal gradients. The accuracy of the model's predictions has been rigorously validated against independent measurements, achieving an accuracy within 12% of these external data points. This level of accuracy substantiates the model's effectiveness and supports its application in predicting geothermal gradient trends in regions that are underexplored or entirely unexplored due to the absence of direct measurements.

The application of machine learning techniques, specifically the Gradient-Boosted Regression Tree algorithm, has proved instrumental in synthesizing complex data into understandable and actionable insights. This approach has allowed us to generate a comprehensive geothermal gradient map of Colombia, which serves as a crucial tool in identifying regions with significant geothermal potential. The map serves as indicative for areas with elevated geothermal gradients, suggesting promising locations for future exploration and resource assessment.

The findings from this study can be particularly impactful for the geothermal energy sector in Colombia. The identified high-potential areas can guide future exploration efforts, ensuring that resources are allocated efficiently and effectively to locations most likely to yield substantial geothermal resources. This strategic focus is expected to significantly advance Colombia's capabilities in geothermal energy production, supporting the country's goals for sustainable energy development and contributing to global renewable energy efforts.

Looking forward, the study underscores the need for continued enhancements in machine learning methodologies to further refine the accuracy of geothermal gradient predictions. It is recommended that future research incorporate more localized and detailed geological data, which could provide deeper insights into subsurface conditions and improve model predictions. Additionally, exploring new geophysical features that might influence geothermal gradients could unveil further intricacies of geothermal systems.

Moreover, as the geothermal gradient data used in this study are derived from average measurements across varied depths, it is crucial to approach the interpretation of these data with an understanding of their inherent limitations. The integration of direct, high-resolution temperature and depth measurements could significantly enhance the accuracy of future predictions and provide a more detailed understanding of geothermal systems.

This research demonstrates the efficacy of Machine Learning tools for estimating geothermal gradients, particularly in the challenging context of Colombia. The accuracy and practicality of the model validate the value of data-driven approaches in enhancing exploration techniques with scarcity of direct measurements. Nevertheless, it is acknowledged that the algorithm has considerable scope for further improvement, particularly in the handling of the complexity of geothermal data and the enhancement of prediction accuracy across diverse geological settings. This work provides a robust foundation for the advancement of these techniques, with ongoing refinements anticipated to enhance the precision and effectiveness of these predictive models in geothermal resource assessment.

## CRediT authorship contribution statement

**Juan C. Mejía-Fragoso:** Writing – review & editing, Writing – original draft, Visualization, Validation, Resources, Methodology, Investigation, Formal analysis, Data curation, Conceptualization. **Manuel A. Flórez:** Writing – review & editing, Supervision, Methodology, Investigation. **Rocío Bernal-Olaya:** Writing – review & editing, Supervision, Investigation, Conceptualization.

## Declaration of competing interest

The authors declare that they have no known competing financial interests or personal relationships that could have appeared to influence the work reported in this paper.

## Data availability

Data supporting the findings of this study are available in the GitHub repository associated with this project. Interested parties can access the dataset, additional resources, and the code used for the study at the following URL: https://github.com/jcmefra/Geothermal-Gradient-Machine-Learning The complete data set of apparent geothermal gradient can be requested from the Servicio Geológico Colombiano (SGC) via their online assistance channel: https://www.sgc.gov.co/.

## Acknowledgments

We thank Servicio Geológico Colombiano (SGC) for generously collecting and providing the data used in this study. Manuel A. Florez received financial support from Vicerrectoría de Investigación y Extensión - Universidad Industrial de Santander, under Grant No. 3755.

## Appendix A. Supplementary data

Supplementary material related to this article can be found online at https://doi.org/10.1016/j.geothermics.2024.103074.